\documentclass[%
reprint,
superscriptaddress,
 amsmath,amssymb,
 aps,
pra,
]{revtex4-2}

\usepackage{graphicx}
\usepackage{dcolumn}
\usepackage{bm}
\usepackage{hyperref}
\usepackage{braket}
\usepackage{mathtools}
\usepackage{amsthm}
\usepackage{xcolor}
\usepackage{caption}
\usepackage{algorithm} 
\usepackage{algorithmic}
\usepackage{ragged2e}
\usepackage{makecell}
\usepackage{multirow}

\theoremstyle{definition}

\begin{document}

\preprint{APS/123-QED}

\title{Two Variations of Quantum Phase Estimation for Reducing Circuit Error Rates: Application to the Harrow--Hassidim--Lloyd Algorithm}

\author{Yonghae Lee} \email{Contact author: yonghaelee@kangwon.ac.kr}
\affiliation{
Department of Liberal Studies, Kangwon National University, Samcheok 25913, Republic of Korea}

\author{Minjin Choi} 
\affiliation{
Center for Quantum Information R\&D, Korea Institute of Science and Technology Information, Daejeon 34141, Republic of Korea}

\author{Youngho Min} 
\affiliation{
Ingenium College of Liberal Arts, Kwangwoon University, Seoul 01897, Republic of Korea}

\author{Eunok Bae} 
\affiliation{
Electronics and Telecommunications Research Institute, Daejeon 34129, Korea}

\author{Sunghyun Bae} 
\affiliation{Department of AI convergence Electronic Engineering, Sejong University, 209, Neungdong-ro, Gwangjin-gu, Seoul 05006, Republic of Korea}

\date{\today}

\begin{abstract}
We introduce two variations of the quantum phase estimation algorithm: quantum \textit{shifted} phase estimation and quantum \textit{punctured} phase estimation. The shifted method employs a bit-string left shift to discard the most significant bit and focus on lower-order phase components, and the punctured method removes qubits corresponding to known phase bits, thereby streamlining the circuit. To demonstrate the effectiveness of the two variations, we integrate them into a hybrid quantum-classical implementation of the Harrow--Hassidim--Lloyd algorithm for solving linear systems. The hybrid method leverages both quantum and classical processors to identify and remove unnecessary qubits and gates. As a result, our method reduces qubit and gate counts compared to previous implementations, leading to lower overall circuit error rates on current hardware. Experimental demonstrations on IBM superconducting hardware confirm the error-mitigation effectiveness of the proposed hybrid method.
\end{abstract}

\maketitle


\section{Introduction}

Quantum phase estimation (QPE)~\cite{Kitaev1995} is a fundamental quantum algorithm for accurately determining the eigenvalues of a unitary operator. It serves as a key subroutine in a wide range of applications~\cite{Brassard2002,Giovannetti2006,AspuruGuzik2005,Lloyd2014}, from quantum simulation~\cite{Lloyd1996} and integer factoring~\cite{Shor1994} to the Harrow--Hassidim--Lloyd (HHL) algorithm~\cite{Harrow2009}. Notably, Shor's algorithm~\cite{Shor1994} factors large integers in polynomial time, far exceeding the capabilities of the best known classical algorithms and thereby threatening the security of RSA cryptosystems~\cite{Rivest1978}. The HHL algorithm~\cite{Harrow2009} provides an exponential speedup over existing classical methods for solving certain systems of linear equations. 

The versatility of the QPE algorithm stems from its ability to output an $n$-bit binary approximation of the phase, yielding successive bits from the first through the $n$th. However, in more general scenarios the original QPE algorithm can prove inefficient. For example, a user who wishes to bypass the first 100 bits and begin estimation at a later position still incurs the full cost of preparing all preceding qubits. In another case, if certain bits of the phase are already known, it would be wasteful to re-estimate those bits rather than focusing solely on the remaining unknown portion. To address these practical use-cases, we introduce two variations of the QPE algorithm designed for a range of settings.

We address the first scenario by incorporating the concept of a \textit{bit‐string left shift}~\cite{Cormen2009}. By shifting the entire $n$-bit estimate one position to the left, the most significant bit is discarded. The remaining bits move into higher-order positions. As a result, the new bit pattern represents the lower-order portion of the phase. We designate this procedure the quantum shifted phase estimation (QSPE) algorithm. The second scenario admits a more intuitive modification of the QPE circuit. Based on the pre-known phase bits, one can identify the corresponding qubits and remove them from the register. One then adapts the controlled-unitary gates and the inverse quantum Fourier transform~\cite{Nielsen2010} to operate on the reduced set of qubits. Because this approach \textit{punctures} the circuit at the positions of known bits, we refer to it as the quantum punctured phase estimation (QPPE) algorithm.

The first part of this paper is devoted to the design and exposition of the two algorithms, QSPE and QPPE. In the second part, we demonstrate how these methods can be applied to reduce circuit error rates in the HHL algorithm.

In order for quantum algorithms to function as intended and achieve quantum advantage~\cite{Harrow2017,Preskill2018,Zhong2021}, several assumptions on qubits, gates, and measurements are required~\cite{Aaronson2015}. Significant efforts are underway to develop quantum computers on diverse physical platforms~\cite{Devoret2013,Blatt2012,OBrien2007,Saffman2010}. However, fully overcoming hardware‐induced errors remains elusive. In practice, when quantum algorithms are executed on current devices, gate, measurement, and decoherence errors accumulate throughout the circuit, resulting in an elevated overall circuit error rate~\cite{Preskill2018,Temme2017}. Such error accumulation constitutes a critical bottleneck that degrades quantum algorithm performance.

Among the various error‐mitigation strategies in the literature~\cite{Fowler2012,Temme2017,Kandala2019,Huggins2021}, one approach is to replace a segment of the quantum algorithm with classical computation~\cite{Lee2019}. By offloading classically tractable subroutines, the remaining quantum circuit can be executed with fewer qubits and gates, leading to a lower overall circuit error rate.

As a second contribution of this work, we integrate QSPE and QPPE into the HHL algorithm to construct a \textit{hybrid} quantum–classical HHL method. We first identify eigenvalue information for a system of linear equations using the QPE algorithm, and then employ classical processors to reduce the numbers of qubits and gates required in the HHL circuit. Our hybrid HHL algorithm constitutes the most general form of this paradigm~\cite{Lee2019,Zhang2022}, achieving significantly reduced qubit and gate counts for solving linear systems. The resulting reduction in circuit complexity directly translates into diminished error accumulation and enhanced algorithmic performance.

The main contributions of this work are as follows.
\begin{itemize}
 \item QSPE: Extension of QPE via a bit-string left shift.
 \item QPPE: Extension of QPE via puncturing qubits corresponding to prior phase information.
 \item Development of theoretical foundations for phase-information utilization, including binary matrices, minimal distinguishing column sets, and a taxonomy of phase-estimation qubit types.
 \item Design of a hybrid HHL algorithm with reduced circuit error rate, achieving highly efficient use of qubits and gates.
\end{itemize}

The remainder of this paper is organized as follows. Section~\ref{sec:QPE} reviews the original QPE algorithm~\cite{Kitaev1995} and its gate-level operation, providing foundational background for the bit-string left shift and prior-information concepts introduced in later sections. Section~\ref{sec:observations} then examines two modifications to the 4-qubit QPE circuit: the addition of a controlled-$U$ gate and the removal of unnecessary qubits. Section~\ref{sec:results} presents our QSPE and QPPE variations of QPE. To demonstrate the application of these variations, Section~\ref{sec:OverviewHHL} revisits the HHL algorithm, and Section~\ref{sec:PIforHHL} discusses eigenvalue extraction from the linear system $A\vec{x} = \vec{b}$ and the classification of phase-estimation qubits. Section~\ref{sec:Application} presents the hybrid HHL algorithm and describes the circuit construction for a concrete example. A comparison of its qubit and gate requirements with those of previous methods is also provided. Lastly, Section~\ref{sec:Experiments} demonstrates via experiments on IBM hardware that our hybrid method substantially reduces circuit error rates.

\section{Circuit for QPE Algorithm} \label{sec:QPE}

The QPE algorithm~\cite{Kitaev1995} extracts the eigenphase of a unitary operator when supplied with its eigenstate. To derive two variations of QPE, we begin by reviewing the quantum Fourier transform, its inverse, and the QPE circuit.

\subsection{Quantum Fourier Transform} \label{sec:QFT}

The quantum Fourier transform (QFT) is the quantum analogue of the discrete Fourier transform~\cite{Nielsen2010}. It acts on an $n$-qubit register by mapping each computational‐basis state $\ket{x}$ to a uniform superposition with phases proportional to $x$. On the orthonormal basis $\{\ket{0}, \ket{1}, \dots, \ket{2^n - 1}\}$, QFT is defined as the linear operator
\begin{equation}
\ket{j} \xrightarrow{\mathrm{QFT}} \frac{1}{\sqrt{2^n}} \sum_{k=0}^{2^n-1} e^{2\pi i \frac{jk}{2^n}} \ket{k}.
\end{equation}
It is often convenient to express $\ket{j}$ in its binary form $j = (j_1 j_2 \dots j_n)_2$. With this notation, the action of QFT can also be written in a product representation:
\begin{eqnarray}
&&\ket{j_1}\otimes\ket{j_2}\otimes\cdots\otimes\ket{j_n} \nonumber \\
&&\xrightarrow{\mathrm{QFT}} \left( \frac{\ket{0}+e^{2\pi i 0.j_n}\ket{1}}{\sqrt{2}} \right) \otimes \left( \frac{\ket{0}+e^{2\pi i 0.j_{n-1}j_n}\ket{1}}{\sqrt{2}} \right) \nonumber \\
&& \quad \qquad \otimes \cdots \otimes \left( \frac{\ket{0}+e^{2\pi i 0.j_1j_2\ldots j_n}\ket{1}}{\sqrt{2}} \right). \label{eq:PR_QFT}
\end{eqnarray}

\begin{figure}
\includegraphics[clip,width=.99\columnwidth]{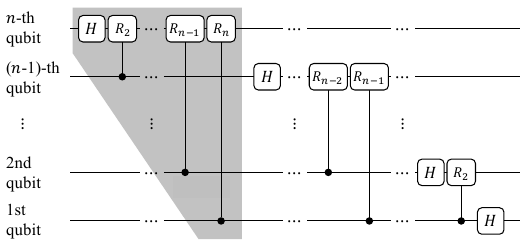}
\caption{\justifying Circuit for the $n$-qubit QFT. This circuit comprises only Hadamard and controlled-$R_k$ gates. The shaded region highlights the sequence of a Hadamard gate followed by controlled-$R_k$ gates that encodes the input labels $j_1,\dots,j_n$ into the phase of the target qubit. The final layer of SWAP gates, which effects the qubit reversal, has been omitted. In this figure, the bottom wire corresponds to the first qubit and the top wire to the last, which reverses the usual top-to-bottom numbering. The same convention applies throughout.
}
\label{fig:QFT}
\end{figure}

From this representation, the circuit for QFT can be implemented using the Hadamard gate $H$ and the phase gate $R_k$, defined as
\begin{equation}
H = \frac{1}{\sqrt{2}}
\begin{bmatrix}
1 & 1\\
1 & -1
\end{bmatrix},
\qquad
R_k = 
\begin{bmatrix}
1 & 0\\
0 & e^{\frac{2\pi i}{2^k}}
\end{bmatrix}.
\end{equation}
Fig.~\ref{fig:QFT} shows the circuit for the $n$-qubit QFT without the SWAP gates, and our results depend critically on the roles of the Hadamard gate and the controlled-$R_k$ gates in this circuit.

To see how the qubit labels become encoded into the phase of a single target qubit, focus on the shaded region in Fig.~\ref{fig:QFT}. First, the Hadamard gate acting on the target qubit $\ket{j_{1}}$ encodes $j_1$ as the most significant phase bit of that target. Next, the controlled-$R_2,R_3,\dots,R_n$ gates are applied sequentially. For each $k = 2,\ldots,n$, the controlled-$R_k$ gate imparts the phase contribution $j_k$ onto the $k$-th phase bit of the same target qubit. Consequently, the change induced by the shaded region can be expressed as follows:
\begin{eqnarray}
&&\ket{j_1}\otimes\ket{j_2}\otimes\cdots\otimes\ket{j_n} \nonumber \\
&&\xrightarrow{~~~} \left( \frac{\ket{0}+e^{2\pi i 0.j_1j_2\ldots j_n}\ket{1}}{\sqrt{2}} \right) \otimes\ket{j_2}\otimes\cdots\otimes\ket{j_n}. \label{eq:1Shaded}
\end{eqnarray}
To achieve this behavior for every possible binary value of $j$, the circuit applies controlled operations throughout.

As shown in Fig.~\ref{fig:QFT}, if the SWAP gates are omitted from the $n$-qubit QFT circuit, then for each qubit $j$, the remaining circuit consists of a single Hadamard gate followed by $j-1$ controlled-$R_k$ rotations~\cite{Nielsen2010}. Hence, the total number of Hadamard and controlled-rotation gates required in Fig.~\ref{fig:QFT} is
\begin{equation}
\frac{n(n+1)}{2}. \label{eq:numberQFT}
\end{equation}

\subsection{Inverse Quantum Fourier Transform} \label{sec:IQFT}

The inverse quantum Fourier transform (IQFT) reverses the action of QFT, converting a phase-encoded superposition on an $n$-qubit register back into its computational-basis amplitudes. In the context of the QPE algorithm, we review the circuit implementation of IQFT.

\begin{algorithm}[H]
\caption{$n$-qubit IQFT$'$}
\label{alg:IQFTp}
\begin{algorithmic}[1]
\REQUIRE An $n$-qubit register in the phase-encoded product state given by Eq.~\eqref{eq:IQFTp}
\ENSURE The same $n$-qubit register in the computational-basis state $\ket{j_n}\otimes\cdots\otimes\ket{j_1}$
\STATE Apply a Hadamard gate $H$ to the last qubit.
\FOR{$k = n-1,n-2,\dots,1$}
\FOR{$l = n,n-1,\dots,k+1$}
\STATE Apply the controlled-$R^\dagger_{l-k+1}$ gate with control on qubit $l$ and target on qubit $k$.
\ENDFOR
\STATE Apply a Hadamard gate $H$ to qubit $k$.
\ENDFOR
\end{algorithmic}
\end{algorithm}

Since IQFT is the inverse of QFT, its circuit can be directly obtained from the QFT circuit. In practice, a layer of SWAP gates is appended to the end of the QFT circuit. Accordingly, in the IQFT circuit, these SWAP gates appear at the beginning. To reduce the overall gate count in the QPE circuit, we employ a modified version of IQFT in which all SWAP gates are omitted. We denote this modification by IQFT$'$. Since SWAP gates only reverse the qubit order, their removal reverses the order of the output qubits of IQFT$'$. Hence, IQFT$'$ acts as follows:
\begin{eqnarray}
&& \left( \frac{\ket{0}+e^{2\pi i0.j_n}\ket{1}}{\sqrt{2}} \right) \otimes \cdots \otimes \left( \frac{\ket{0}+e^{2\pi i0.j_1j_2\ldots j_n}\ket{1}}{\sqrt{2}} \right) \nonumber \\
&&\xrightarrow{\mathrm{IQFT}'} \ket{j_n}\otimes\cdots\otimes\ket{j_1}, \label{eq:IQFTp}
\end{eqnarray}
which yields Algorithm~\ref{alg:IQFTp}. The number of Hadamard and controlled-rotation gates required to realize this operation in the circuit matches the value specified in Eq. Eq.~\eqref{eq:numberQFT}.

\begin{figure}
\includegraphics[clip,width=.99\columnwidth]{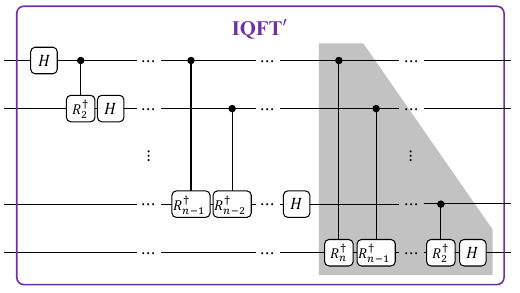}
\caption{\justifying Circuit for the $n$-qubit IQFT$'$. This circuit is obtained by omitting the SWAP gates originally appended to the original IQFT circuit. As a result, the qubit ordering is reversed, and the assignment of subsequent gates to each qubit is correspondingly adjusted.
}
\label{fig:IQFTp}
\end{figure}

Fig.~\ref{fig:IQFTp} illustrates the circuit for the $n$-qubit IQFT$'$. We examine the shaded region in Fig.~\ref{fig:IQFTp}, whose action is expressed as
\begin{eqnarray}
&&\ket{j_n}\otimes\cdots\otimes\ket{j_2}\otimes\left( \frac{\ket{0}+e^{2\pi i 0.j_1j_2\ldots j_n}\ket{1}}{\sqrt{2}} \right) \nonumber \\
&&\xrightarrow{~~~} \ket{j_n}\otimes\cdots\otimes\ket{j_2}\otimes\ket{j_1}. \label{eq:2Shaded}
\end{eqnarray}
To transfer the phase bit $j_1$ into the qubit label, the bits $j_n$ through $j_2$ are sequentially removed using controlled-$R_k^\dagger$ gates.

\subsection{QPE Algorithm} \label{sec:QPEA}

The QPE algorithm is a fundamental subroutine in quantum computing that estimates the phase $\varphi$ of an eigenvalue $e^{2\pi i \varphi}$ of a unitary operator $U$, given access to its corresponding eigenvector. To be precise, assume $U$ has an eigenvector $\ket{u}$ with eigenvalue $e^{2\pi i \varphi}$, where $\varphi$ is unknown. The goal of the QPE algorithm~\cite{Kitaev1995,Nielsen2010} is to estimate $\varphi$ in its binary expansion to a desired precision.

The circuit implementation of the QPE algorithm employs two quantum registers, $F$ and $G$. Register $F$ comprises $n$ qubits, all initialized to $\ket{0}$, i.e. $\ket{0}^{\otimes n}$. Here, $n$ is chosen based on the required number of bits of accuracy in the estimate of $\varphi$. Register $G$ holds the eigenstate $\ket{u}$ and contains as many qubits as are needed to represent it.

In the QPE circuit, the $n$ qubits of register $F$ serve as control qubits and register $G$ supplies the target eigenstate. Specifically, for each $k=1,2,\dots,n$, a controlled-$U^{2^{k-1}}$ gate uses the $k$-th qubit of $F$ as control and acts on the state in $G$, imprinting a phase $2\pi ( 2^{k-1}\varphi)$ onto the $k$-th qubit of the control register $F$. As a result, the successive binary digits of $\varphi$ become encoded across the qubits of $F$. To implement the circuit in this way, the QPE algorithm uses a black box (or oracle) to apply controlled-$U$ operations, without requiring direct access to or knowledge of the unitary operator $U$.

\begin{algorithm}[H]
\caption{$n$-qubit QPE algorithm}
\label{alg:QPEA}
\begin{algorithmic}[1]
\REQUIRE A black box for $U$ that implements controlled-$U^{2^k}$ operations
\REQUIRE A register $G$ prepared in the eigenstate $\ket{u}$ of $U$ with eigenvalue $e^{2\pi i \varphi}$

\ENSURE An $n$-bit estimate $\widetilde{\varphi} = (0.\varphi_1 \varphi_2 \dots \varphi_n)_2$ of the phase $\varphi$

\STATE Initialize an $n$-qubit register $F$ in $\ket{0}^{\otimes n}$ and $G$ in $\ket{u}$.
\STATE Apply a Hadamard gate $H$ to each qubit of register $F$.
\FOR{$k = 1,2,\dots,n$}
\STATE Apply the controlled-$U^{2^{k-1}}$ operation with control on qubit $k$ of $F$ and target register $G$.
\ENDFOR
\STATE Apply $n$-qubit IQFT$'$ to register $F$.
\STATE Measure each qubit of register $F$ in the computational basis to obtain bits $\varphi_1, \varphi_2, \dots, \varphi_n$.
\end{algorithmic}
\end{algorithm}

For an $n$-qubit register $F$, the QPE circuit can be implemented in the steps of Algorithm~\ref{alg:QPEA}. It employs IQFT$'$ instead of the original IQFT. As noted above, IQFT circuit includes SWAP gates, and since their effect is fully understood, this modification causes no problems.

\begin{figure}
\includegraphics[clip,width=.99\columnwidth]{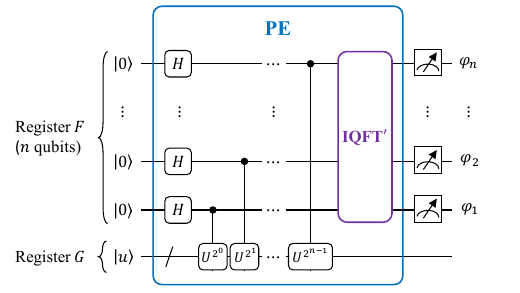}
\caption{\justifying Circuit of the QPE algorithm. The input state $\ket{0}^{\otimes n}\otimes\ket{u}$ is prepared on registers $F$ and $G$. For each $k$, a controlled-$U^{2^{k-1}}$ gate is applied to register $G$ using the $k$-th qubit of $F$ as the control. Register $F$ is then measured to extract the $n$-bit estimate $\widetilde{\varphi}=0.\varphi_1\varphi_2\cdots\varphi_n$ of the phase $\varphi$. The phase-estimation subcircuit (PE step) is used as a subroutine in the HHL algorithm.
}
\label{fig:QPE}
\end{figure}

After applying all controlled-$U^{2^{k-1}}$ gates, the state of register $F$ becomes
\begin{eqnarray}
&&\left( \frac{\ket{0}+e^{2\pi i 0.\varphi_n}\ket{1}}{\sqrt{2}} \right) \otimes \left( \frac{\ket{0}+e^{2\pi i 0.\varphi_{n-1}\varphi_n}\ket{1}}{\sqrt{2}} \right) \nonumber \\
&&\otimes \cdots \otimes \left( \frac{\ket{0}+e^{2\pi i 0.\varphi_1\varphi_2\ldots \varphi_n}\ket{1}}{\sqrt{2}} \right),
\end{eqnarray}
where $\varphi_j$ is the $j$-th bit in the binary expansion of $\varphi$. When IQFT$'$ is applied and its action is given by Eq.~\eqref{eq:IQFTp}, register $F$ holds an estimate of $\varphi$'s binary digits. Fig.~\ref{fig:QPE} illustrates the circuit for the QPE algorithm using IQFT$'$.

To implement the $n$-qubit QPE circuit, the phase register $R$ is first initialized by applying a Hadamard gate to each of its $n$ qubits, requiring $n$ Hadamard gates in total. During the phase-encoding stage, one applies controlled–$U^{2^{k-1}}$ for $k=1,\dots,n$, which in the black-box model corresponds to
\begin{equation}
\sum_{k=1}^n2^{k-1}=2^n-1
\end{equation}
queries to the black box~\cite{Cleve1998}. Finally, IQFT$'$ requires a total of $\tfrac{n(n+1)}{2}$ gates, comprising Hadamard gates and controlled-$R_k$ rotations.

In practice, the number $t$ of qubits in register $F$ must be chosen to achieve both $n$-bit precision and a desired success probability~\cite{Nielsen2010}. To be precise, if one wishes to estimate $\varphi$ to within $\pm 2^{-n}$ with probability at least $1 - \varepsilon$, then the number of qubits required in register $F$ is
\begin{equation}
t = n + \left\lceil \log_{2}\left(2 + \tfrac{1}{2\varepsilon}\right)\right\rceil. \label{eq:precision}
\end{equation}
The extra $t-n$ qubits serve to suppress the probability of a large estimation error below $\varepsilon$. In other words, using $t$ qubits guarantees that the output $\widetilde{\varphi}$ satisfies $\lvert \varphi - \widetilde{\varphi} \rvert < 2^{-n}$ with probability at least $1 - \varepsilon$.

\section{Observations} \label{sec:observations}

In this section, we modify the QPE circuit in two distinct ways and observe how each modification affects the extractable phase information. These observations motivate the main ideas behind the two QPE variations introduced in the next section. To build intuition, we consider the 4-qubit QPE algorithm. Suppose a unitary operator $U$ has an eigenstate $\ket{u}$ whose phase is to be estimated with four‐bit precision; that is, $\widetilde{\varphi} = (0.\varphi_{1}\varphi_{2}\varphi_{3}\varphi_{4})_2$.

\subsection{Observation 1: Left-Shifting Circuit for Phase Bits} \label{sec:observation1}

\begin{figure*}
\centering
\includegraphics[width=\textwidth]{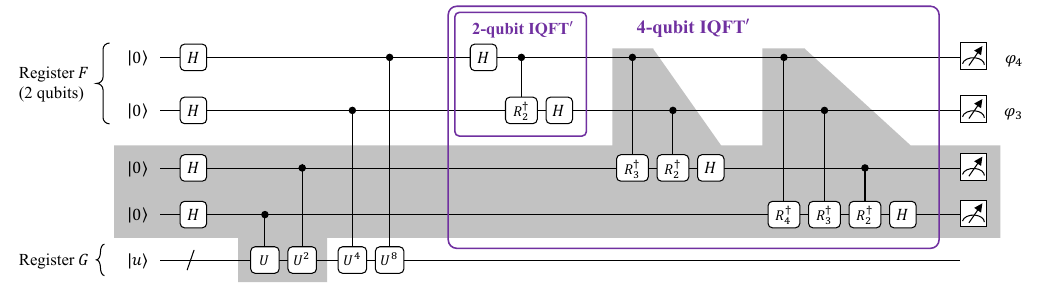}
\caption{\justifying
Left-shifting circuit for phase bits. This circuit is derived by omitting the shaded region from the original 4-qubit QPE circuit. More precisely, the first and second qubits of register $F$ are unused, and all associated gates and measurements are removed. The third and fourth bits of the phase estimate $\widetilde{\varphi}$ are obtained by executing the circuit on the remaining third and fourth qubits and measuring them. This example demonstrates that the deletion of the leading qubits of register $F$ induces a left-shifting operation on the phase bit string.
} \label{fig:Observation1}
\end{figure*}

For the first observation, we remove the first and second qubits of register $F$ in the 4‐qubit QPE circuit, along with all gates and measurements connected to those wires. In Fig.~\ref{fig:Observation1}, the shaded region indicates the portions of the 4‐qubit QPE circuit that are removed. As a result, only two qubits of register $F$ remain, so we expect to obtain a two‐bit estimate of the phase.

Concretely, after applying the controlled‐$U^{8}$ gate in the modified circuit of Fig.~\ref{fig:Observation1}, the remaining (third and fourth) qubits of register $F$ are in the state
\begin{equation}
\left(\frac{\ket{0} + e^{2\pi i0.\varphi_{4}}\ket{1}}{\sqrt{2}}\right)_{\mathrm{4th}}
 \otimes 
\left(\frac{\ket{0} + e^{2\pi i0.\varphi_{3}\varphi_{4}}\ket{1}}{\sqrt{2}}\right)_{\mathrm{3rd}}.
\end{equation}
It is interesting to note that the deletion of the wires (and associated gates) for the first and second qubits causes the 4-qubit IQFT$'$ to reduce to a 2-qubit IQFT$'$. This 2-qubit IQFT$'$ transfers the first two phase bits it receives into the computational-basis labels. Therefore, after executing IQFT$'$, the third and fourth qubits become
\begin{equation}
\ket{\varphi_{4}}_{\mathrm{4th}} \otimes \ket{\varphi_{3}}_{\mathrm{3rd}}.
\end{equation}
By measuring each of these two qubits, one can estimate the third and fourth bits of the phase $\varphi$.

From this observation, we learn that whereas the original QPE algorithm always estimates the phase bits from $\varphi_{1}$ through $\varphi_{n}$ in sequence, our modified circuit makes it possible to skip the first two qubits and directly estimate the third and fourth bits. This behavior is analogous to a left-shift operation in bit manipulations. A comparison between the 2-qubit QPE circuit that estimates bits $\varphi_{1}, \varphi_{2}$ and the modified circuit of Fig.~\ref{fig:Observation1} that estimates bits $\varphi_{3}, \varphi_{4}$ reveals that the latter requires more controlled-$U^k$ gates. Hence, the additional controlled unitary gates can be viewed as performing a left shift on the phase information.

\subsection{Observation 2: Punctured Circuit for Known Phase Bits} \label{sec:observation2}

Unlike in the first observation, the second observation requires prior knowledge of certain bits of the phase. Suppose that the second and fourth bits of the phase $\varphi$ are known to be 0 and 1, respectively. In other words, the estimate takes the form $\widetilde{\varphi} = (0.\varphi_{1}0\varphi_{3}1)_2$. Intuitively, since two out of the four phase bits are already known, we only need to estimate the remaining two bits. These unknown bits, $\varphi_{1}$ and $\varphi_{3}$, correspond to the first and third qubits of register $F$, so we anticipate that only those two qubits are needed for estimation.

\begin{figure*}
\centering
\includegraphics[width=\textwidth]{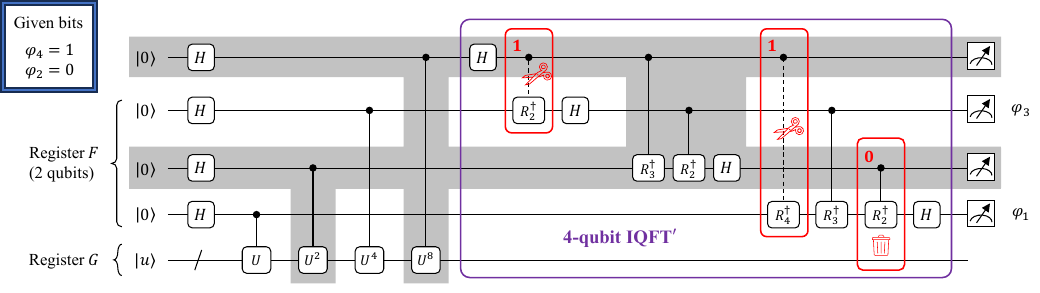}
\caption{\justifying
Punctured circuit for known phase bits. This circuit is derived from the original 4-qubit QPE circuit under the assumption of prior knowledge of certain phase bits. In this example, we assume $\varphi_2 = 0$ and $\varphi_4 = 1$. Based on this information, the second and fourth qubits of register $F$ are rendered inactive, and all associated gates and measurements are removed. The original circuit contains gates acting on the first and third qubits, which are controlled by the second and fourth qubits. These gates are then modified accordingly. The first and third bits of the phase estimate $\widetilde{\varphi}$ are obtained by executing the circuit on the remaining first and third qubits and measuring them. This example demonstrates how puncturing the circuit to remove known-qubit operations enables phase estimation over the remaining unknown bits.
} \label{fig:Observation2}
\end{figure*}

To verify this intuition, let us modify the 4-qubit QPE circuit again. As shown in Fig.~\ref{fig:Observation2}, we remove the second and fourth qubits of register $F$ in the 4-qubit QPE circuit, along with all gates and measurements connected to those wires. Specifically, all gates targeting the second and fourth qubits are deleted, and controlled-$R^\dagger_{k}$ gates that use them as control qubits are either applied unconditionally or omitted, depending on the known bit values. In Fig.~\ref{fig:Observation2}, all wires and gates that are removed are indicated by shaded regions, while any circuit elements that are modified (rather than deleted) are enclosed by unshaded outlines.

Next, we examine how the states of the first and third qubits of register $F$ evolve step by step. After applying the controlled-$U^{4}$ gate in the circuit of Fig.~\ref{fig:Observation2}, the remaining qubits of register $F$ are in the state
\begin{equation}
\left(\frac{\ket{0} + e^{2\pi i0.\varphi_{3}1}\ket{1}}{\sqrt{2}}\right)_{\mathrm{3rd}} \otimes \left(\frac{\ket{0} + e^{2\pi i0.\varphi_{1}0\varphi_{3}1}\ket{1}}{\sqrt{2}}\right)_{\mathrm{1st}}.
\end{equation}
To transfer the estimated bits into the computational-basis labels of the first and third qubits, we apply the modified 4-qubit IQFT$'$ depicted in Fig.~\ref{fig:Observation2}. Since the fourth bit $\varphi_{4}$ is known to be 1, the controlled-$R^\dagger_{2}$ gate on the third qubit can be replaced by a single-qubit $R^\dagger_{2}$. Similarly, since the first qubit's phase contains the bit string $\varphi_{1}0\varphi_{3}1$, we apply $R^\dagger_{4}$ to the first qubit and omit the third controlled-$R^\dagger_{2}$ gate as it is unnecessary. As a result, after executing this modified IQFT$'$, the first and third qubits become
\begin{equation}
\ket{\varphi_3}_{\mathrm{3rd}} \otimes \ket{\varphi_1}_{\mathrm{1st}}.
\end{equation}
Measuring each of these two qubits then yields the first and third bits of the phase $\varphi$.

From this observation, we can draw the following conclusions. First, if certain bits of the phase are known in advance, it is possible to estimate only the remaining unknown bits. Specifically, if the $k$-th bit of the phase is known, then the $k$-th qubit of register $F$ is not required for estimation. This situation can be likened to a punctured register in which gaps appear at the positions of known bits. Furthermore, Fig.~\ref{fig:Observation2} depicts a punctured circuit corresponding to the case where two nonconsecutive bits are known; from this example, one can infer that the same approach applies to the general case in which known bits are adjacent or when more than two bits are known.

\section{Results} \label{sec:results}

Based on the observations in Section~\ref{sec:observations}, we introduce two variations of the original QPE algorithm~\cite{Kitaev1995}. The first algorithm allows one to estimate only the phase bits corresponding to a user-specified bit interval. This can be regarded as applying the concept of left-shifting the binary representation of the phase during computation. The second algorithm exploits prior knowledge of certain phase bits. Since specific bits are known in advance, it is more efficient to apply the algorithm only to the remaining, unknown bits. In practice, one removes the qubits and gates corresponding to the known bits, thereby performing phase estimation solely on the unknown subset. This concept can be likened to a punctured arrangement of qubits in the circuit.

\subsection{Result 1: Quantum Shifted Phase Estimation Algorithm}

We devise the QSPE algorithm, a variant of the QPE algorithm that incorporates left bit shifting to enable the estimation of specific intervals of phase information.

The QPE algorithm estimates the phase $\varphi$ in binary by obtaining its bits sequentially, starting from the most significant bit; that is, $\widetilde{\varphi}=(0.\varphi_{1}\varphi_{2}\cdots\varphi_{n})_2$. In contrast, QSPE applies a left shift of $s$ bits to $\varphi$ and then extracts the next $n$ bits at once, thus yielding an estimate of $\widetilde{\varphi}=(0.\varphi_{s+1}\varphi_{s+2}\cdots\varphi_{s+n})_2$.

\begin{figure}
\includegraphics[clip,width=.99\columnwidth]{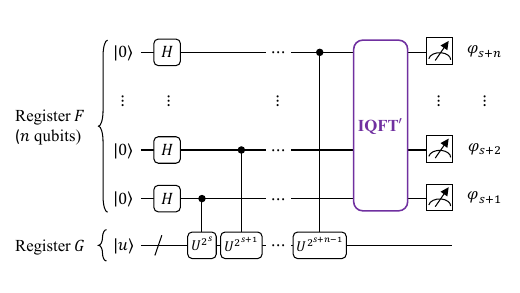}
\caption{\justifying Circuit of the QSPE algorithm. The input state $\ket{0}^{\otimes n}\otimes\ket{u}$ is prepared in registers $F$ and $G$. 
For a non‐negative integer $s$ and each $k\in\{1,2,\dots,n\}$, a controlled‐$U^{2^{s+k-1}}$ gate acts on register $G$, using the $k$-th qubit of $F$ as control. After applying $n$-qubit IQFT$'$, register $F$ is measured to obtain the $n$-bit estimate $\widetilde{\varphi} = (0.\varphi_{s+1}\varphi_{s+2}\cdots\varphi_{s+n})_2$ of the phase $\varphi$.
}
\label{fig:QSPE}
\end{figure}

Specifically, we describe the procedure of the QSPE algorithm and explain how the input state evolves throughout the algorithm. The circuit of the QSPE algorithm requires two registers, $F$ and $G$. Register $F$ is initialized in $\ket{0}^{\otimes n}$ and $G$ in $\ket{u}$, as shown in Fig.~\ref{fig:QSPE}. A Hadamard gate $H$ on each qubit of $F$ produces
\begin{equation}
\frac{1}{\sqrt{2^n}} \sum_{y=0}^{2^n-1} \ket{y}_F \otimes \ket{u}_G.
\end{equation}
For a non‐negative integer $s$ and for each $k\in\{1,2,\dots,n\}$, a controlled-$U^{2^{s+k-1}}$ operation is applied, using the $k$-th qubit of $F$ as control, so that the state becomes
\begin{equation}
\frac{1}{\sqrt{2^n}} \sum_{y=0}^{2^n-1} e^{2\pi i (2^s\varphi)y}\ket{y}_F \otimes \ket{u}_G.
\end{equation}
Applying $n$-qubit IQFT$'$ to register $F$ then yields
\begin{equation}
\frac{1}{2^n}\sum_{x=0}^{2^n-1} \sum_{y=0}^{2^n-1} e^{2\pi iy\left((2^s\varphi) - \tfrac{x}{2^n}\right)} \ket{x}_F \otimes \ket{u}_G. \label{eq:QSPE_U}
\end{equation}

We now examine the meaning of $2^s\varphi$. The value $2^s\varphi$ can be decomposed as
\begin{equation}
2^s\varphi = \varphi_\text{int} + \varphi_\text{frac},
\end{equation}
where the \textit{integer} part $\varphi_\text{int}$ and the \textit{fractional} part $\varphi_\text{frac}$ of $2^s\varphi$ are given by
\begin{eqnarray}
\varphi_\text{int} &=& (\varphi_1 \varphi_2 \cdots \varphi_s)_2, \\
\varphi_\text{frac} &=& (0.\varphi_{s+1}\cdots \varphi_{s+n} \cdots)_2.
\end{eqnarray}
Because the integer part does not affect the phase of the eigenvalue, we have
\begin{equation}
e^{2\pi i y \left((2^s\varphi) - \frac{x}{2^n}\right)}
= e^{2\pi i y \left(\varphi_\text{frac} - \frac{x}{2^n}\right)}.
\end{equation}
As a result, \eqref{eq:QSPE_U} can be represented as
\begin{equation}
\frac{1}{2^n}\sum_{x=0}^{2^n-1} \sum_{y=0}^{2^n-1} e^{2\pi i y \left(\varphi_\text{frac} - \frac{x}{2^n}\right)} \ket{x}_F \otimes \ket{u}_G.
\end{equation}

By measuring the qubits of register $F$, we obtain the $n$-bit estimate
\begin{equation}
\widetilde{\varphi}=(0.\varphi_{s+1}\varphi_{s+2}\cdots\varphi_{s+n})_2,
\end{equation}
which provides the estimates for the first $n$ bits of the fractional part $\varphi_\mathrm{frac}$. The insertion of the controlled-$U^{2^s}$ operation thus shifts the bit string left by $s$ positions. Consequently, the QSPE algorithm yields phase estimates for bits $s+1$ through $s+n$.

\begin{algorithm}[H]
\caption{$n$-qubit QSPE algorithm}
\label{alg:QSPEA}
\begin{algorithmic}[1]
\REQUIRE A black box for $U$ that implements controlled-$U^{2^k}$ operations 
\REQUIRE A register $G$ prepared in the eigenstate $\ket{u}$ of $U$ with eigenvalue $e^{2\pi i \varphi}$ 
\REQUIRE A non‐negative integer $s$ specifying the left shift 
\ENSURE An $n$‐bit estimate $\widetilde{\varphi} = (0.\varphi_{s+1}\varphi_{s+2}\dots\varphi_{s+n})_2$ of the phase $\varphi$ 

\STATE Initialize register $F$ in $\ket{0}^{\otimes n}$ and register $G$ in $\ket{u}$. 
\STATE Apply a Hadamard gate $H$ to each qubit of register $F$. 
\FOR{$k = 1,2,\dots,n$}
\STATE Apply the controlled-$U^{2^{s + k - 1}}$ operation with control on qubit $k$ of $F$ and target register $G$. 
\ENDFOR 
\STATE Apply $n$-qubit IQFT$'$ to register $F$. 
\STATE Measure each qubit of register $F$ in the computational basis to obtain bits $\varphi_{s+1}, \varphi_{s+2}, \dots, \varphi_{s+n}$. 
\end{algorithmic}
\end{algorithm}

Algorithm \ref{alg:QSPEA} illustrates the implementation of the QSPE circuit using an $n$-qubit register $F$. In QSPE, the integer parameter $s$ determines only which contiguous block of $n$ bits in the binary expansion of the phase $\varphi$ is to be estimated; it does not enter into the determination of how many total qubits $t$ are required in register $F$. Indeed, since QSPE performs a original QPE on the fractional part of $2^{s}\varphi$, the same success‐probability analysis applies. In particular, if one wishes to guarantee that the $n$-bit interval $\varphi_{s+1}\varphi_{s+2}\cdots\varphi_{s+n}$ is estimated to within $\pm 2^{-n}$ with probability at least $1-\varepsilon$, then register $F$ must contain $t$ qubits, regardless of the chosen shift $s$, where $t$ is given by Eq.~\eqref{eq:precision}. Thus, while $s$ affects only which bits of $\varphi$ are revealed, it does not increase or decrease the total number of qubits required to achieve a given $n$-bit precision and failure probability $\varepsilon$.

Let us now quantify the resources required by the $n$-qubit QSPE circuit in terms of gates and black-box queries. The only modification relative to the original QPE circuit is the introduction of the integer shift parameter $s$. This adjustment impacts solely the number of black-box queries: left-shifting the phase by $s$ to obtain finer resolution incurs extra queries. Concretely, during the phase-encoding stage an additional controlled–$U^{2^s}$ is applied on each of the $n$ control qubits. Hence the total number of queries to the black box becomes
\begin{equation}
\left(\sum_{k=1}^n2^{k-1} \right) + 2^s n. 
\end{equation}
The shift has no effect on the number of Hadamard gates or controlled-$R_k$ rotations required for state preparation or for IQFT$'$.

\subsection{Result 2: Quantum Punctured Phase Estimation Algorithm}

Another variation of the QPE algorithm is the QPPE algorithm. Unlike QSPE, QPPE requires prior information about certain bits of the phase. In this case, the sequential estimation of the phase from the first bit onward (as in original QPE) can be inefficient when some bits are already known. We therefore devise a procedure that directly estimates only the unknown bits.

Consider a register $F$ composed of $n$ qubits, used to estimate the phase $\varphi$ to $n$-bit precision. To formalize the prior information, let 
\begin{equation}
P \subseteq \{1,2,\dots,n\}
\end{equation}
be the index set of bit positions whose values are assumed known. If $k \in P$, then the $k$-th bit of the binary expansion of $\varphi$ is assumed known. Recall that the $k$-th qubit is normally used to estimate the $k$-th bit of the phase; this correspondence is illustrated in Fig.~\ref{fig:QPE}. Hence, it is reasonable to exclude the $k$-th qubit from the estimation whenever $k \in P$.

In Section~\ref{sec:observation2}, we confirmed by a specific example that this approach is feasible. More generally, for any index set $P$, the same strategy applies. To show this, we analyze the QPE circuit in reverse. First, recall from Section~\ref{sec:IQFT} how IQFT$'$ decodes phase bits into qubit labels. IQFT$'$ then uses controlled-rotation gates to remove those bits. If certain bits are known a priori, the corresponding controlled-rotation gates can be omitted or implemented unconditionally. Hence, prior bit information permits the omission of qubits that would otherwise serve as controls in the IQFT$'$ stage.

If we know which qubits are not used in IQFT$'$, their initial state becomes irrelevant. In other words, there is no need to encode phase information onto those qubits. Therefore, we can forgo preparing them in superposition via Hadamard gates and skip applying controlled‐$U$ operations to them. As a result, prior knowledge of certain phase bits enables a reduction in the number of qubits and gates needed to implement the QPE circuit, thereby simplifying the hardware requirements.

Fortunately, the ordering of the binary digits obtained by phase estimation aligns with the physical layout of qubits in register $F$ as connected by circuit wires, making this simplification intuitive. Consider the QPE circuit in which IQFT$'$ is used. If prior information is provided for specific bits of the phase $\varphi$, we simply remove the corresponding qubits and their wires from the circuit. Complementing this, the IQFT$'$ subcircuit is adjusted to reflect the known bits. Because this process resembles punching a horizontal hole through the circuit, we refer to it as the \textit{punctured} version.

\begin{algorithm}[H]
\caption{$n$-qubit QPPE algorithm}
\label{alg:QPPEA}
\begin{algorithmic}[1]
\REQUIRE A black box for $U$ that implements controlled-$U^{2^k}$ operations 
\REQUIRE A register $G$ prepared in the eigenstate $\ket{u}$ of $U$ with eigenvalue $e^{2\pi i \varphi}$ 
\REQUIRE An index set $P \subseteq \{1,2,\dots,n\}$ specifying the positions of the known bits, i.e., for each $p\in P$, the value of $\varphi_{p}$ is known
\ENSURE An ($n-|P|$)-bit estimate of the phase $\varphi$; the $n$-bit estimate is punctured at the positions of the known bits

\STATE Initialize register $F$ in $\ket{0}^{\otimes n}$ and register $G$ in $\ket{u}$.

\FOR{$k = 1,2,\dots,n$}
 \IF{$k \notin P$}
\STATE Apply a Hadamard gate $H$ to qubit $k$ of $F$.
\STATE Apply the controlled-$U^{2^{k - 1}}$ operation with control on qubit $k$ of $F$ and target register $G$.
 \ENDIF
\ENDFOR 

\IF{$n \notin P$}
\STATE Apply a Hadamard gate $H$ to the last qubit of $F$.
\ENDIF

\FOR{$k = n-1,n-2,\dots,1$}
 \IF{$k \notin P$}
\FOR{$l = n,n-1,\dots,k+1$}
 \IF{$l \in P$ \AND $\varphi_{l} = 1$}
\STATE Apply the $R^\dagger_{l-k+1}$ gate to qubit $k$ of $F$.
 \ELSIF{$l \notin P$}
\STATE Apply the controlled-$R^\dagger_{l-k+1}$ gate with control on qubit $l$ and target on qubit $k$.
 \ENDIF
\ENDFOR
\STATE Apply a Hadamard gate $H$ to qubit $k$ of $F$.
 \ENDIF
\ENDFOR

\STATE Measure those qubits of register $F$ that are not in $P$ to obtain an ($n-|P|$)-bit estimate of the phase $\varphi$.
\end{algorithmic}
\end{algorithm}

Algorithm \ref{alg:QPPEA} illustrates the implementation of the QPPE circuit using an $n$-qubit register $F$. In the algorithm, not all $n$ qubits are used for phase estimation. If an index set $P$ of size $|P|$ is given, meaning that $|P|$ bits of the phase are known in advance, then only $n - |P|$ qubits in register $F$ are actually required. Although the pseudocode above is written as if all $n$ qubits are present for the sake of intuition, in practice the number of qubits scales down in direct proportion to $|P|$.

How many qubits must register $F$ contain in order to guarantee high‐confidence estimation via the QPPE algorithm? Although QPPE targets irregularly‐spaced bits, it ultimately produces a bit string of length $n - |P|$, and what matters is how closely this $(n - |P|)$-bit estimate approximates the true phase. Hence, from the perspective of measurement outcomes, QPE, QSPE, and QPPE all share the same formula given in Eq.~\eqref{eq:precision}. Specifically, if one wants to estimate the unknown $(n - |P|)$-bit portion of the phase $\varphi$ to within $\pm2^{-(n - |P|)}$ with probability at least $1 - \varepsilon$, then register $F$ must contain the following number of qubits:
\begin{equation}
\left(n - \lvert P\rvert\right) + \left\lceil \log_{2}\left(2 + \tfrac{1}{2\varepsilon}\right)\right\rceil.
\end{equation}

In this section, we have introduced two variations of the QPE algorithm. Note that if no left shift is applied in the first method, or if no prior bit information is available in the second method, each reduces to the original QPE algorithm. In this sense, both approaches extend the QPE framework by incorporating additional flexibility. Our algorithms enable flexible phase estimation using a register $F$ with a limited number of qubits. In the following sections, we describe how both methods can be applied within the HHL algorithm.

\section{Overview of HHL Algorithm and Circuit Implementation} \label{sec:OverviewHHL}

The second topic of this work is to demonstrate that the application of QSPE and QPPE to the HHL algorithm can reduce its circuit error rate. To this end, we introduce the HHL algorithm and examine the principles underlying its operation in this section.

\subsection{HHL Algorithm}

For a given linear system $A\vec{x} = \vec{b}$, the HHL algorithm provides an approximation of the expectation value $\vec{x}^\dagger M \vec{x}$ for an operator $M$~\cite{Harrow2009}. Throughout this work, we assume that $A$ is Hermitian and has been pre-scaled such that its eigenvalue spectrum lies within the interval $(0,1)$

\begin{figure}
\includegraphics[clip,width=.99\columnwidth]{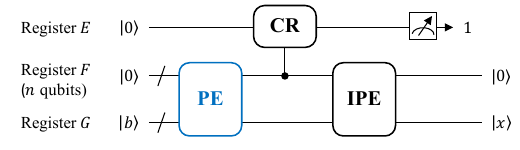}
\caption{\justifying HHL algorithm. The HHL algorithm consists of three steps: PE, CR, and IPE, which are implemented using three registers. The state $ \ket{0} \otimes \ket{0}^{\otimes n} \otimes \ket{b} $ is prepared in these registers as the input for the algorithm. The PE step corresponds to the QPE algorithm without measurement. The CR step is implemented via a controlled rotation operation, whose effect is described in Eq.~\eqref{eq:CRoutput}. The IPE step is the inverse of the PE step. Upon measuring 1 in Register $E$, the quantum state in registers $F$ and $G$ collapses to $\ket{0}^{\otimes n} \otimes \ket{x}$.
}
\label{fig:HHL}
\end{figure}

The HHL algorithm can be divided into three main steps, excluding state preparation and measurement.
The three sequential steps of the HHL algorithm are the phase estimation (PE) step, the controlled rotation (CR) step, and the inverse phase estimation (IPE) step. These steps are illustrated in Fig.~\ref{fig:HHL}. To implement the HHL circuit, we employ three quantum registers, denoted $E$, $F$, and $G$. Register $E$, consisting of a single qubit, is used in the CR step. Register $F$, comprising $n$ qubits, is employed for estimating the eigenvalues of the matrix $A$. Register $G$ encodes the input vector $\vec{b}$ in quantum form.

\begin{algorithm}[H]
\caption{$n$-qubit HHL algorithm}
\label{alg:HHL}
\begin{algorithmic}[1]
\REQUIRE A Hermitian matrix $A$ encoded as a unitary $U = e^{2\pi i A}$ 
\REQUIRE An input state $\ket{b}_G$ in register $G$ 
\ENSURE An approximate solution state $\ket{x}\propto A^{-1}\ket{b}$ in $G$ 

\STATE Initialize register $F$ in $\ket{0}^{\otimes n}$, register $E$ in $\ket{0}$, and register $G$ in $\ket{b}$. 
\STATE Apply Hadamard gates $H^{\otimes n}$ to $F$. 
\FOR{$k = 1,\dots,n$}
\STATE Apply controlled-$U^{2^{k-1}}$ with control on the $k$-th qubit of $F$ and target $G$. 
\ENDFOR 
\STATE Apply $n$-qubit IQFT$'$ on $F$. 

\STATE For each basis state $\ket{j}_F$ of register $F$ with estimated eigenvalue $\widetilde\lambda_j$, apply a rotation $R_y\bigl(2\arcsin(c/\widetilde\lambda_j)\bigr)$ on $E$ controlled by $\ket{j}_F$. 

\STATE Apply the inverse IQFT$'$ on $F$. 
\FOR{$k = n,\dots,1$}
\STATE Apply controlled-$U^{-2^{k-1}}$ with control on the $k$-th qubit of $F$ and target $G$. 
\ENDFOR 
\STATE Apply Hadamard gates $H^{\otimes n}$ to $F$ to reset it to $\ket{0}^{\otimes n}$. 

\STATE Measure $E$. If outcome is $1$, the state of $G$ is proportional to $A^{-1}\ket{b}$; otherwise, the algorithm fails and should be repeated. 
\end{algorithmic}
\end{algorithm}

Algorithm~\ref{alg:HHL} presents the procedure for the $n$-qubit HHL algorithm. In particular, we begin with the input state
\begin{equation}
\ket{0}_E \otimes \ket{0}_F \otimes \ket{b}_G,
\end{equation}
where $\ket{0}_F = \ket{0}^{\otimes n}$ and $\ket{b}_G$ encodes the input vector $\vec{b}$. For the eigenbasis $\{\ket{u_j}\}$ of $A$, the state $\ket{b}_G$ can be expanded as
\begin{equation}
\ket{b}_G = \sum_{j=1}^{m} \alpha_j \ket{u_j}_G,
\end{equation}
where the coefficients $\alpha_j$ satisfy the normalization condition.

The PE step corresponds to the QPE algorithm without measurement, where the unitary used in the controlled‐$U^{2^k}$ gates is $U = e^{2\pi i A}$. Unlike the original QPE algorithm, in the HHL algorithm the matrix $A$ and the vector $\vec{b}$ are given explicitly, so no black‐box oracle is required. Accordingly, upon completion of the PE step, register $F$ holds an estimate of each eigenvalue $\lambda_j$. This process transforms the input state as follows:
\begin{eqnarray}
&&\ket{0}_F \otimes \sum_{j=1}^{m} \alpha_j \ket{u_j}_G \nonumber \\
&&\xrightarrow{\mathrm{PE}} \sum_{j=1}^{m} \sum_{x,y=0}^{2^n-1} \frac{1}{2^n} e^{2\pi i y \left( \lambda_j - \frac{x}{2^n} \right)} \ket{x}_F \otimes \alpha_j \ket{u_j}_G.
\end{eqnarray}
For simplicity, we consider an ideal case where each eigenvalue $\lambda_j$ of $A$ can be represented with $n$ bits, i.e.,
\begin{equation}
\lambda_j = \left(0.b_{j,1}b_{j,2}\dots b_{j,n}\right)_2.
\end{equation}
Thus, for each eigenvalue $\lambda_j$, its estimate $\widetilde{\lambda}_j$ is identical to $\lambda_j$. In this idealized scenario, the output state of the PE step becomes
\begin{equation}
\sum_{j=1}^{m} \alpha_j \ket{b_{j,1}b_{j,2}\dots b_{j,n}}_F \otimes \ket{u_j}_G. \label{eq:PE}
\end{equation}

In the CR step, we perform controlled rotations on the register $E$ conditioned on the eigenvalue estimates Eq.~\eqref{eq:PE} in register $F$. The post-rotation state becomes
\begin{equation}
\sum_{j=1}^{m} \left( \sqrt{1-\frac{c^2}{\widetilde{\lambda}_j^2}}\ket{0}_E + \frac{c}{\widetilde{\lambda}_j}\ket{1}_E \right)
\otimes \alpha_j \ket{2^n\widetilde{\lambda}_j}_F \otimes \ket{u_j}_G, \label{eq:CRoutput}
\end{equation}
where $c = O(1/\kappa)$ and $\kappa$ is the condition number of $A$~\cite{Harrow2009}. This transformation is implemented by applying a controlled-$R_y(2\arcsin(c/\widetilde{\lambda}_j))$ gate on register $E$ with register $F$ as the control. The single-qubit rotation $R_y(\theta)$ about the $y$-axis has the matrix form 
\begin{equation}
R_y(\theta)
=
\begin{bmatrix}
\cos\tfrac{\theta}{2} & -\sin\tfrac{\theta}{2} \\
\sin\tfrac{\theta}{2} & \cos\tfrac{\theta}{2}
\end{bmatrix}.
\end{equation}

In the IPE step, we apply the inverse phase estimation to return register $F$ to the initial state $\ket{0}_F$. When we measure Register $E$ and obtain the outcome 1, register $G$ collapses to
\begin{equation}
\sum_{j=1}^m \frac{\alpha_j}{\widetilde{\lambda}_j} \ket{u_j}_G,
\end{equation}
which corresponds to the normalized solution of the linear system $A\vec{x} = \vec{b}$. In this work, we focus on this normalized solution rather than the expectation value $\vec{x}^\dagger M \vec{x}$, in order to compare the performance of different circuit implementations of the HHL algorithm.

\subsection{Circuit Implementation of HHL Algorithm} \label{sec:CircuitHHL}

In this work, we describe how the QSPE and QPPE methods can be applied to the HHL algorithm and show that this approach leads to a reduction in the circuit error rate. In this subsection, to enable a reasonable comparison with previous results, we clarify the circuit implementation of each step of the HHL algorithm.

For the PE step, an implementation of the QPE algorithm exists~\cite{Nielsen2010}; we adopt the basic circuit shown in Fig.~\ref{fig:QPE} (with the PE block highlighted), which employs IQFT without SWAP gates to reduce the total gate count. When the PE step is implemented in this way, the overall state in \eqref{eq:PE} becomes
\begin{equation}
\sum_{j=1}^{m} \alpha_j \ket{b_{j,n}b_{j,n-1}\dots b_{j,1}}_F \otimes \ket{u_j}_G. \label{eq:PEp}
\end{equation}
This expression shows that the ordering of qubits in register $F$ directly corresponds to the bit positions of the estimated eigenvalues: the $k$-th qubit holds the $k$-th bit $b_{j,k}$ of $\lambda_j$.

We now calculate the number of gates required in the PE step. This step can be divided into three parts.
(i) In the first part, a Hadamard gate is applied to each of the $n$ qubits to prepare a maximally entangled state. This requires $n$ Hadamard gates.
(ii) Next, as explained in Section~\ref{sec:QPEA}, the $n$-qubit QPE algorithm requires $2^n - 1$ black-box queries to implement the controlled-$U^{2^{k-1}}$ operations. In contrast, the HHL algorithm employs Hamiltonian simulation~\cite{Berry2007,Childs2010} to realize the unitary matrix $U = e^{2\pi i A}$. More precisely, for a given time parameter $t$, Hamiltonian simulation yields a decomposition of $e^{itA}$ into a sequence of elementary gates (e.g., single-qubit gates and CNOT gates). Let $h(A)$ denote the number of elementary gates required to implement the controlled-$e^{itA}$ operation. By choosing $t = 2^k \pi$, one can implement the controlled-$U^{2^{k-1}}$ operation using $h(A)$ elementary gates. Hence, a total of $nh(A)$ elementary gates are required for this part.
(iii) Finally, the number of Hadamard and controlled-$R_k^\dagger$ gates in IQFT$'$ is $\tfrac{n(n+1)}{2}$.

For the CR step, no general scheme is known for implementing it directly using only elementary gates. This is because the authors of~\cite{Harrow2009}, who first introduced the HHL algorithm and analyzed its complexity, did not provide a concrete circuit implementation. Although the CR operation can be described using logical gates, its decomposition into elementary gates tends to be complicated and may require a large number of gates. Therefore, instead of constructing a detailed CR circuit, we focus on analyzing the number of qubits that must serve as control qubits. Since the HHL algorithm does not have access to the eigenvalue information at runtime, the CR step must, in principle, treat all $n$ qubits of register $F$ as control qubits. This requirement imposes a considerable gate overhead in implementing the CR step.

Finally, because the IPE step is simply the inverse of the PE step, its circuit implementation follows directly from that of the PE step. As a result, aside from the CR step, the total gate count of the HHL circuit is determined by the implementation of the PE step; hence, in this work we substitute an analysis of the PE step for a separate treatment of the IPE step.

\section{Prior Information for HHL Algorithm} \label{sec:PIforHHL}

The HHL algorithm internally relies on the eigenvalues and eigenstates of the matrix $A$. In particular, it dedicates a substantial number of qubits and gates to estimating these eigenvalues and computing their reciprocals. When prior knowledge of the eigenvalues is available, one can reduce the qubit and gate overhead in the circuit implementation of the HHL algorithm. This idea lies at the heart of the hybrid method~\cite{Lee2019}. In this section, we generalize the idea by introducing the binary matrix of eigenvalue estimates.

\subsection{QPE algorithm and binary matrix} \label{sec:binary}

We construct a binary matrix of eigenvalue estimates by repeatedly applying the QPE algorithm, and introduce notation for its columns under specified conditions.

As in the hybrid method~\cite{Lee2019}, we run the QPE algorithm to gather eigenvalue information for the matrix $A$. We employ the $n$-qubit QPE circuit shown in Fig.~\ref{fig:QPE}, using the unitary $U = e^{2\pi i A}$. In each run, we sequentially measure the qubits of register $F$ from the first through the $n$-th to populate one row of a binary matrix with entries in $\{0,1\}$, where each row represents a distinct eigenvalue estimate. We repeat this procedure until no new rows appear. The binary matrix has $n$ columns, which equals the number of qubits in register $F$.

Through this process, we can obtain an $m \times n$ binary matrix $B$:
\begin{equation}
B = 
\begin{bmatrix}
b_{11} & b_{12} & \cdots & b_{1n} \\
b_{21} & b_{22} & \cdots & b_{2n} \\
\vdots & \vdots & \ddots & \vdots \\
b_{m1} & b_{m2} & \cdots & b_{mn}
\end{bmatrix}.
\end{equation}
Once the binary matrix $B$ is obtained, a classical computer is used to identify a subset of columns that can distinguish all rows. Specifically, let $D \subseteq \{1,2,\dots,n\}$ be a (minimal) distinguishing column set for $B$, i.e., for any $j \neq k$,
\begin{equation}
B_{j,D} \neq B_{k,D},
\end{equation}
where $B_{j,D}$ denotes the sub-row of the $j$th row restricted to columns in $D$. Given the matrix $B$ and the distinguishing column set $D$, we classify the columns of $B$ as follows:
\begin{itemize}
 \item Columns in $D$ are called \textit{distinguishing columns}. The one with the smallest index is the \textit{leading column}.
 \item Among the remaining columns, those with all entries identical are \textit{constant columns}.
 \item The rest are \textit{non-constant columns}.
\end{itemize}

For illustration, consider the $3\times6$ binary matrix
\begin{equation}
B =
\begin{bmatrix}
1 & 1 & 0 & 1 & 0 & 1 \\
0 & 1 & 1 & 1 & 0 & 0 \\
0 & 1 & 0 & 0 & 0 & 1
\end{bmatrix}. \label{eq:firstB}
\end{equation}
In general, a minimal distinguishing column set is not unique. For the matrix $B$, we choose $D = \{3,4\}$ as one minimal distinguishing set, since each pair of rows differs with respect to these columns. The smallest index in $D$, namely the third column, is the leading column. The second and fifth columns are constant, consisting of all ones and all zeros, respectively. The first and sixth columns are non-constant but non-distinguishing: their entries vary across rows, yet they are not required by $D$ to distinguish every row pair.

The problem of finding a minimal distinguishing column set in a binary matrix goes by various names in different fields. It is a combinatorial optimization problem known to be NP-hard~\cite{Garey1979}. For small matrices, exhaustive (brute-force) search can be applied~\cite{Cormen2009}, whereas for larger instances, greedy heuristics~\cite{Chvatal1979} or integer linear programming formulations~\cite{Bertsimas2005} may be employed to obtain approximate or exact solutions.

\subsection{Distinguishing and Non-distinguishing Qubits and Their Effects} \label{eq:effects}

The binary matrix encodes some of the eigenvalue information required at each step of the HHL algorithm. Using this matrix, we can reduce the number of qubits and gates in the HHL circuit.

We first review how the qubits of register $F$, the binary representation of eigenvalues, and the columns of $B$ align. To be specific, the $j$-th qubit of $F$ encodes the $j$-th eigenvalue bit, which corresponds to the $j$-th column of the binary matrix $B$. Accordingly, we classify the qubits of $F$ into three types:
\begin{itemize}
 \item Distinguishing qubits (the one with the smallest index is the leading qubit).
 \item Non-distinguishing constant qubits.
 \item Non-distinguishing non-constant qubits.
\end{itemize}
Each qubit's type follows from the type of its corresponding column in $B$.

\begin{table}
\caption{\justifying Effects of qubit types on the HHL circuit. The binary matrix $B$ permits classification of the qubits in register $F$ by type. Certain qubit types can be omitted from one or more steps of the HHL algorithm. Those qubits that are unnecessary across all three algorithmic steps may be removed from the full HHL circuit.
}
\label{table:summary}
\begin{tabular}{c||c|c|c||l}
\hline
Type& PE & CR & IPE & \makecell[l]{Full \\ circuit} \\
\hline \hline
Distinguishing qubits & Keep & Keep & Keep & Keep \\
\hline
\makecell[l]{Constant qubits following \\ the leading qubit} & Omit & Omit & Omit & Remove \\
\hline
\makecell[l]{Non-constant qubits \\ following the leading qubit} & Keep & Omit & Keep & Keep \\
\hline
\makecell[l]{Non-distinguishing qubits \\ preceding the leading qubit} & Omit & Omit & Omit & Remove \\
\hline
\end{tabular}
\end{table}

We then examine how each qubit type influences the steps of the HHL algorithm, as shown in Table~\ref{table:summary}. In the CR step, distinguishing qubits serve as the control qubits, while non‐distinguishing qubits remain inactive. Suppose that among the $n$ qubits of register $F$, $t$ qubits are non-distinguishing. This means that $t$ columns of the binary matrix are unnecessary for distinguishing all eigenvalue estimates, only the remaining $n-t$ distinguishing columns are required. In the original HHL algorithm, since these distinctions are unknown at runtime, all $n$ qubits must serve as control qubits, and the CR step must handle $2^n$ input states. However, if one knows in advance which $t$ qubits are non-distinguishing, one can restrict control to the $n-t$ distinguishing qubits and reduce the number of input states to at most $2^{n-t}$.

In the PE step, only the bits corresponding to distinguishing qubits are estimated to enable their role as controls in the CR step. Consequently, phase estimation need only run from the leading qubit through the last distinguishing qubit. For non‐distinguishing qubits, whether they are estimated depends on their position relative to the leading qubit. (i) Any qubit preceding the leading qubit is automatically non‐distinguishing and is not used to estimate its corresponding bit. Therefore, phase estimation is performed only on bits from the leading qubit onward, which can be implemented using a QSPE circuit. (ii) Following the leading qubit, constant qubits require no estimation, as their bit values are fixed and known from the binary matrix $B$. By using a QPPE circuit, one can omit phase estimation for these qubits. (iii) Last, let us consider the non-constant qubits located after the leading qubit. Unlike constant qubits, their estimated bits take the values 0 or 1 probabilistically; hence, they cannot be excluded by the QPPE circuit. The core principle of QPPE is that when certain bits are known with certainty, the corresponding controlled-rotation gates can be omitted or implemented unconditionally.

Note that the omission of non-distinguishing gates in the PE step does not affect the subsequent CR step. This is because the entries of non-distinguishing columns are known \textit{a priori} from the binary matrix $B$. These known values can be encoded directly into the controlled-rotation angles during the CR step.

In summary, the PE step requires only the distinguishing qubits and the non-constant qubits following the leading qubit, whereas the CR step relies exclusively on the distinguishing qubits. The effect of each qubit type on the IPE step mirrors its role in the PE step. Table~\ref{table:summary} summarizes the types of qubits that can be omitted at each step, enabling one to identify which qubits may be removed from the full circuit.

\section{Application: Hybrid HHL algorithm} \label{sec:Application}

How can QSPE and QPPE be applied to reduce the circuit error rate of quantum algorithms? In this section, we devise a \textit{hybrid} HHL algorithm that incorporates these methods. We then describe how to apply this hybrid approach to a concrete linear system. Finally, we compare the resulting reductions in qubit and gate counts with those of previous methods.

\subsection{Hybrid HHL algorithm}

\begin{figure*}
\centering
\includegraphics[width=\textwidth]{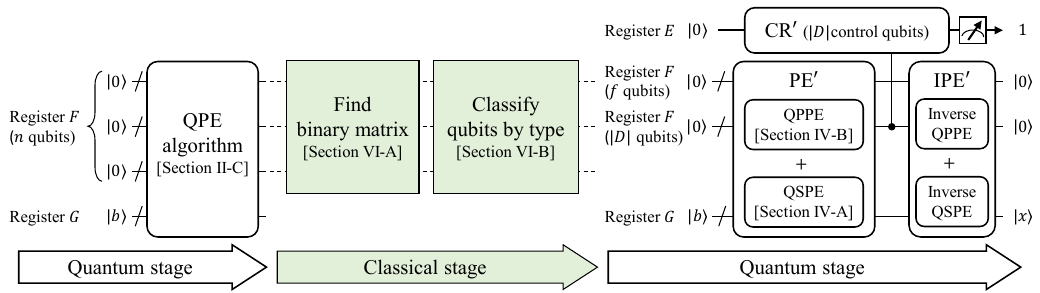}
\caption{\justifying
Hybrid HHL algorithm. Given the linear system $A\vec{x} = \vec{b}$, we repeatedly apply the QPE algorithm on the unitary $U = e^{2\pi i A}$ using an $n$-qubit register $F$ to extract eigenvalue information. Measurement outcomes are collected into a binary matrix $B$, and find a distinguishing column set $D$ which is then analyzed to classify each of the $n$ qubits in $F$ as either distinguishing, constant, or non-constant. Let $f$ denote the number of non-constant qubits that follow the leading distinguishing qubit. We construct the PE$'$, CR$'$, and IPE$'$ circuits as follows. Only $|D| + f$ qubits participate in the PE$'$ step; the remaining $n - (|D| + f)$ qubits are removed via QSPE and QPPE. In the CR$'$ step, the $|D|$ distinguishing qubits serve as control qubits, reducing the controlled-rotation inputs to $2^{|D|}$, with rotation angles determined by the corresponding rows of $B$. Finally, the IPE is applied and register $E$ is measured; an outcome of $\ket{1}$ on $E$ yields the normalized solution state in register $G$.
}\label{fig:hybrid}
\end{figure*}

Our hybrid algorithm employs both quantum and classical processors. First, the original QPE algorithm is executed on a quantum computer to extract the eigenvalue information of the matrix $A$, and then these results are processed on a classical computer. Based on this information, the number of qubits and quantum gates required for the HHL circuit can be reduced. In particular, our QSPE and QPPE algorithms are applied to decrease both the qubit and gate counts. The overall process is illustrated in Fig.~\ref{fig:hybrid}.

The hybrid HHL algorithm employs three registers, denoted by $E$, $F$, and $G$, where register $F$ comprises $n$ qubits. The first step of the hybrid HHL algorithm is to perform the QPE algorithm repeatedly on the unitary operator $ U = e^{2\pi i A} $ and the quantum state $\ket{b}_G$, as described in Section~\ref{sec:binary}. This procedure yields a binary matrix $B$ that encodes the eigenvalue information of $A$. Next, a classical computer analyzes the binary matrix $B$ to identify the distinguishing column set $D$. Based on $B$ and $D$, the qubits in register $F$ are classified into three types (distinguishing qubits, constant qubits, and non-constant qubits), which are defined in Section~\ref{eq:effects}.

Given this classification, the HHL circuit is designed with a reduced number of qubits and gates. To be precise, assume that the size of the binary matrix $B$ is $m\times n$. The distinguishing column set $D$ is the index set of columns required to distinguish the $m$ rows of $B$. Assume that $|D| < n$, and let $l$ be the smallest element of $D$. Then, among the $n$ qubits in register $F$, $|D|$ qubits are distinguishing qubits and the remaining $n - |D|$ qubits are non-distinguishing qubits, with the $l$-th qubit designated as the leading distinguishing qubit.

Now we design the HHL circuit for the linear system $A\vec{x} = \vec{b}$. Since we already possess eigenvalue information of $A$ from the QPE algorithm, we can design both qubits and gates efficiently. For convenience, let $s=l-1$ denote the number of non-distinguishing qubits before the leading distinguishing qubit, and $f$ the number of non-constant qubits following it. Table~\ref{table:summary} summarizes which qubit types may be omitted at each step.

The PE step of the original HHL algorithm combines QSPE and QPPE, and is hereafter referred to as the PE$'$ step. In this modified implementation, we omit the $s$ non-distinguishing qubits preceding the leading qubit, since their eigenvalue bits are already encoded in the binary matrix. The QSPE method is then applied to shift each eigenvalue’s binary representation left by $s$ bits, allowing phase estimation to act only on the remaining qubits. Similarly, we apply the QPPE method to remove the $f$ constant qubits following the leading qubit, as each corresponds to a fixed bit value recorded in the binary matrix. Since QSPE and QPPE operate on disjoint sets of qubits and gates, they can be applied in any order without affecting the outcome. This structural independence enables flexible circuit construction. By combining both methods, we obtain the PE$'$ circuit, which requires only $|D| + f$ qubits in register~$F$.

The circuit for the CR step is constructed to produce the state given in Eq.~\eqref{eq:CRoutput}. In the original HHL algorithm, no prior eigenvalue information is available, so the CR step must accommodate all $2^n$ possible phase estimates; accordingly, each of the $n$ qubits in register $F$ serves as a control bit, as described in Section~\ref{sec:CircuitHHL}. By contrast, our hybrid algorithm uses the binary matrix to identify only the $m$ eigenvalues of interest and the distinguishing qubits needed to distinguish them. We refer to this modified version of the CR step as the CR$'$ step. As a result, in the CR$'$ step only the $d$ distinguishing qubits act as control bits, leaving the remaining qubits untouched.

The remaining procedure is identical to that of the original HHL algorithm. We implement the IPE step as the inverse of the PE$'$ circuit. Finally, we measure register $E$ in the computational basis; upon obtaining the outcome $1$, the normalized solution state is prepared in register $G$. Algorithm \ref{alg:hybridHHL} presents the procedure for the $n$-qubit hybrid HHL algorithm.

\begin{algorithm}[H]
\caption{$n$-qubit hybrid HHL algorithm}
\label{alg:hybridHHL}
\begin{algorithmic}[1]
\REQUIRE A Hermitian matrix $A$ encoded as $U = e^{2\pi i A}$
\REQUIRE An input state $\ket{b}$ in register $G$
\ENSURE An approximate solution state $\ket{x}\propto A^{-1}\ket{b}$ in $G$
 
\STATE Perform the $n$-qubit QPE on $U$ and $\ket{b}$ repeatedly to obtain the binary matrix $B$.
\STATE Find a distinguishing column set $D$ of $B$.
\STATE Classify the qubits of register $F$ into distinguishing, constant, and non-constant qubits; denote qubit $l$ as the leading distinguishing qubit; let $P$ be the set of indices of constant qubits following the leading qubit.
\STATE Initialize register $F$ in $\ket{0}^{\otimes n}$, register $E$ in $\ket{0}$, and register $G$ in $\ket{b}$.
 
\STATE \textbf{PE$'$ step (5–6):} Set $s = l - 1$ and prepare an $(n - s)$-qubit QSPE circuit.
\STATE Apply the QPPE method to the QSPE circuit, puncturing the qubits corresponding to $P$, and then execute it.

\STATE \textbf{CR$'$ step:} Using the $|D|$ distinguishing qubits as controls, apply $R_{y}\bigl(2\arcsin(c/\widetilde\lambda)\bigr)$ on register $E$ for each estimated eigenvalue $\widetilde\lambda$.
 
\STATE \textbf{IPE$'$ step:} Execute the inverse of the PE step circuit.
 
\STATE Measure register $E$. If the outcome is $\ket{1}$, the state in $G$ is proportional to $A^{-1}\ket{b}$; otherwise, repeat the algorithm.
\end{algorithmic}
\end{algorithm}

In the $n$-qubit hybrid HHL algorithm, register $F$ consists of $n$ qubits; however, the PE$'$ step in practice utilizes only a subset of them. Specifically, by iteratively applying the QPE algorithm to obtain eigenvalue estimates, the PE$'$ circuit can be implemented using only $|D| + f$ qubits, where $|D| + f < n$. This reduction in qubit count also results in a corresponding decrease in the number of required gates.

\subsection{Example: Application to a Linear System} \label{sec:EXamAppl}

Let us illustrate the hybrid HHL algorithm on a concrete example. We consider a linear system
\begin{equation}
A\vec{x}=\vec{b}, \label{eq:ConEx}
\end{equation}
where $A$ is an arbitrary Hermitian matrix whose three eigenvalues have the following binary expansions:
\begin{eqnarray}
\lambda_1 &=& \frac{7}{16}= (0.011100)_2, \nonumber \\
\lambda_2 &=& \frac{17}{64}=(0.010001)_2, \label{eq:eigenvalues} \\
\lambda_3 &=& \frac{53}{64}=(0.110101)_2.\nonumber
\end{eqnarray}
For each $j$, denote the corresponding eigenstate by $\ket{u_j}$, so that $A\ket{u_j} = \lambda_j\ket{u_j}$. We assume that the user has no prior information about the eigenvalues and that a unit vector $\vec{b}$ can be expressed as a linear combination of these eigenstates,
\begin{equation}
\vec{b}=\sum_{j=1}^3 \alpha_j \ket{u_j}.
\end{equation}
Then the solution to the linear system $A\vec{x} = \vec{b}$ is given by
\begin{equation}
\vec{x}=\sum_{j=1}^3 \frac{\alpha_j}{\lambda_j} \ket{u_j}. \label{eq:realSolution}
\end{equation}

In the first step of the hybrid HHL algorithm, the user repeatedly executes the $n$-qubit QPE algorithm. By collecting the measurement outcomes, the user can obtain the following $3 \times n$ binary matrix:
\begin{equation}
B =
\begin{bmatrix}
1 & 1 & 0 & 1 & 0 & 1 & 0 & 0 & \cdots & 0 \\
0 & 1 & 1 & 1 & 0 & 0 & 0 & 0 & \cdots & 0 \\
0 & 1 & 0 & 0 & 0 & 1 & 0 & 0 & \cdots & 0
\end{bmatrix}, \label{eq:concrete}
\end{equation}
where each row gives the binary estimate of one eigenvalue, i.e.:
\begin{eqnarray}
\widetilde\lambda_1 &=& (0.01110000\dots0)_2, \\
\widetilde\lambda_2 &=& (0.01000100\dots0)_2, \\
\widetilde\lambda_3 &=& (0.11010100\dots0)_2.
\end{eqnarray}
Note that the first six columns of $B$ coincide with those in \eqref{eq:firstB}. From $B$, the user infers that exactly three eigenvalues of $A$ are relevant for solving the linear system; although their true values are unknown, the estimates obtained via the QPE algorithm satisfy the precision bound in Eq.~\eqref{eq:precision}. Consequently, the user proceeds based on these estimated bits.

Next, let us assume that the user finds a minimal distinguishing column set $D=\{3,4\}$. Then, as described in Section~\ref{sec:binary}, the types of all columns are determined by $D$. Because there is a one‐to‐one correspondence between the columns of the binary matrix and the qubits in register $F$, each qubit inherits the type of its associated column. In this example, the qubit types are as follows:
\begin{itemize}
\item Qubits 7–$n$: constant
\item Qubit 6: non-constant
\item Qubit 5: constant
\item Qubit 4: distinguishing
\item Qubit 3: leading and distinguishing
\item Qubit 2: constant
\item Qubit 1: non-constant
\end{itemize}

\begin{figure*}
\centering
\includegraphics[width=\textwidth]{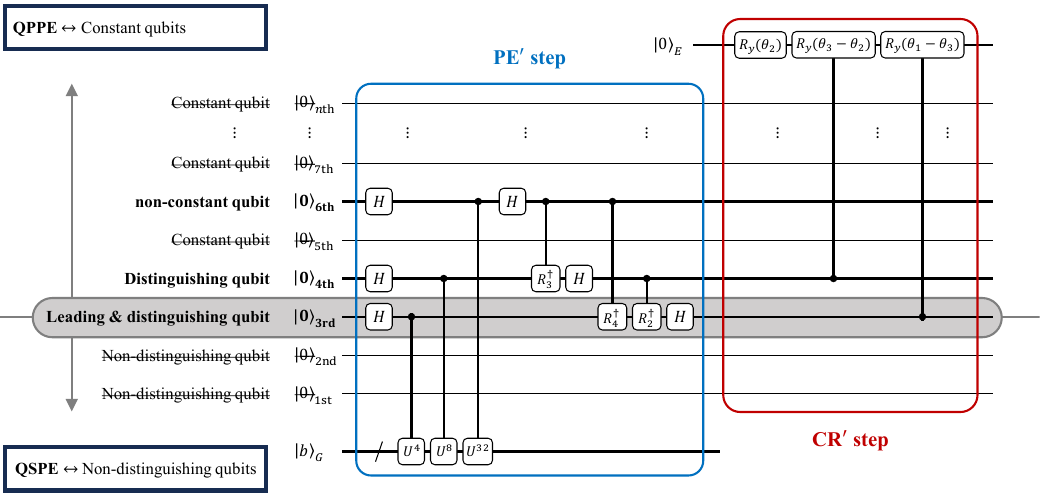}
\caption{\justifying
PE$'$ and CR$'$ circuits of the hybrid HHL algorithm for the concrete example given in Eq.~\eqref{eq:ConEx}.
Assume that register $F$ comprises $n$ qubits and that we solve the linear system whose binary matrix is $B$ from Eq.~\eqref{eq:concrete}. From $B$, the $n$ qubits are classified by type. All non-distinguishing qubits preceding the leading and distinguishing qubit are removed via the QSPE method, and all constant qubits following the leading qubit are removed via the QPPE method. This procedure yields the PE$'$ circuit. In the CR$'$ step, the distinguishing qubits serve as control qubits, whereas the non-distinguishing qubits play no role; this information is then used to construct the CR$'$ circuit. Finally, since the IPE$'$ step reverses the PE$'$ step, the qubits eliminated across all steps are exactly the non-distinguishing qubits before the leading qubit and the constant qubits after it.
} \label{fig:example}
\end{figure*}

Based on the effects of each qubit type presented in Table~\ref{table:summary}, the user prepares the circuits for the PE$'$, CR$'$, and IPE$'$ steps. To prepare the PE$'$ circuit, the user first considers a notional $n$-qubit PE circuit and then eliminates any unnecessary qubits:
(i) According to Table~\ref{table:summary}, all non-distinguishing qubits preceding the leading qubit may be removed from the PE circuit. In this example, the first and second qubits qualify. By applying the QSPE method with a left shift of $s=2$, the first and second qubits can be eliminated.
(ii) Next, all $n-5$ constant qubits following the leading qubit can likewise be removed. In this case, instead of the QSPE method, the user applies the QPPE method to puncture these constant qubits. Because each constant qubit encodes a bit value (0 or 1) during phase estimation, the controlled-rotation gates with constant qubits as controls must be suitably adjusted according to their bit values. Specifically, since all constant qubits represent the value 0, a correspondingly large number of controlled-rotation gates can be further eliminated. Figure~\ref{fig:example} illustrates the PE$'$ circuit after applying QSPE and QPPE.

The user then executes the PE$'$ circuit constructed in this manner, in which eigenvalue estimation is performed using only the third, fourth, and sixth qubits of register $F$. After the PE$'$ step, registers $F$ and $G$ assume the following joint state:
\begin{eqnarray}
& & \alpha_1 \ket{0}_{\mathrm{6th}}\otimes\ket{1}_{\mathrm{4th}}\otimes\ket{1}_{\mathrm{3rd}} \otimes \ket{u_1}_G \nonumber \\
&+& \alpha_2 \ket{1}_{\mathrm{6th}}\otimes\ket{0}_{\mathrm{4th}}\otimes\ket{0}_{\mathrm{3rd}} \otimes \ket{u_2}_G, \nonumber \\
&+& \alpha_3 \ket{1}_{\mathrm{6th}}\otimes\ket{1}_{\mathrm{4th}}\otimes\ket{0}_{\mathrm{3rd}} \otimes \ket{u_3}_G,
\end{eqnarray}
where the $n-3$ remaining qubits of $F$ are omitted.

To prepare the CR$'$ circuit, the user employs the binary matrix $B$ and Table~\ref{table:summary}. The user thereby knows that three quantum states are in superposition at the end of the PE$'$ step and that these states can be distinguished using only those two distinguishing qubits. Thus, the CR$'$ step reduces to using the those two qubits as control qubits to apply the rotation $R_y$ gate on register $E$. Specifically, upon completion of the CR$'$ step, the joint state of the full register should be
\begin{eqnarray}
& & R_y(\theta_1) \otimes \alpha_1 \ket{0}_{\mathrm{6th}}\otimes\ket{11}_{\mathrm{control}} \otimes \ket{u_1}_G \nonumber \\
&+& R_y(\theta_2) \otimes \alpha_2 \ket{1}_{\mathrm{6th}}\otimes\ket{00}_{\mathrm{control}} \otimes \ket{u_2}_G, \nonumber \\
&+& R_y(\theta_3) \otimes \alpha_3 \ket{1}_{\mathrm{6th}}\otimes\ket{10}_{\mathrm{control}} \otimes \ket{u_3}_G,
\end{eqnarray}
where $\theta_j = 2\arcsin(c/\widetilde\lambda_j)$ for each $j$. Figure~\ref{fig:example} shows the CR$'$ circuit implementing this action with one rotation gate and two controlled-rotation gates.

Next, the user performs the IPE$'$ step and subsequent operations. As a result, by employing the hybrid method, only three qubits are used in register $F$ to implement the HHL circuit. Consequently, while a significant number of unnecessary qubits and gates are eliminated, register $G$ holds the normalized version of the solution in Eq.~\eqref{eq:realSolution}, which is identical to the one obtained by the original HHL algorithm.

\subsection{Comparison with previous results} \label{sec:Comparison}

We review previous methods~\cite{Harrow2009,Lee2019,Zhang2022} for implementing the HHL circuit and compare the required number of qubits and gates for the PE$'$, CR$'$, and IPE$'$ steps.

We first refer to the basic method for implementing the original HHL algorithm~\cite{Harrow2009} as \textit{Original09}. For the purpose of comparison with hybrid methods, we assume that although Original09 obtains eigenvalue information via the QPE algorithm, it does not make use of this information. Under this assumption, as described in Section~\ref{sec:CircuitHHL}, the qubit and gate counts for the PE$'$ and IPE$'$ steps can be explicitly quantified. For the CR$'$ step, implementation complexity is characterized by the number of control qubits required for the controlled-rotation gates.

Subsequently, a hybrid method~\cite{Lee2019} was proposed that utilizes eigenvalue information obtained via the QPE algorithm to reduce the number of qubits and gates in the HHL circuit. This method, which we refer to as \textit{Hybrid19}, does not involve notions such as the binary matrix or distinguishing column sets. Instead, it analyzes the measurement outcomes of the QPE algorithm to identify constant qubits and excludes them from the CR step, thereby achieving a substantial reduction in gate count compared to Original09. However, in Hybrid19, the eigenvalue information is used solely to simplify the CR step and is not applied to the PE or IPE steps.

While Zhang \textit{et al}. achieved resource reduction in the PE and IPE steps for a specific $4 \times 4$ linear system~\cite{Zhang2022}, their approach has not been extended to general $n \times n$ cases. We therefore exclude their result from the comparison. The final method considered for comparison is our hybrid approach, which we refer to as \textit{Hybrid25}. As demonstrated in Section~\ref{sec:EXamAppl}, our method reduces the number of qubits and gates required in all three steps of the HHL algorithm.

To compare the qubit and gate counts for each implementation method, we consider the linear system defined in Eq.~\eqref{eq:ConEx}. For Original09, recall that although prior eigenvalue information is obtained, it is not utilized. Accordingly, the basic circuit shown in Figure~\ref{fig:QPE} is adopted to implement its PE$'$ step. Since this circuit serves as a subroutine for phase estimation within the HHL algorithm, the gate counts presented in Section~\ref{sec:CircuitHHL} apply directly. Furthermore, because no prior eigenvalue information is available, the CR$'$ step must account for all possible phase estimates. This implies that all $n$ qubits in register $F$ are used as control qubits in the implementation of the CR$'$ step. Finally, since the IPE$'$ step is the inverse of the PE$'$ step, its gate count is identical. Therefore, the total number of gates for the circuit excluding the CR$'$ step is given by twice the gate count of the PE$'$ step.

In Hybrid19, the user repeatedly executes QPE and stores the eigenvalue estimates on a classical computer. By analyzing these estimates, the user identifies bit positions that are consistent across all estimated values; that is, constant qubits. This information affects only the CR$'$ step, so the PE$'$ and IPE$'$ steps in Hybrid19 remain identical to those in Original09. In contrast, the CR$'$ step excludes the constant qubits, leaving four candidates for control: the first, third, fourth, and sixth qubits in register $F$. Given that QPE reveals three eigenvalue estimates, the CR$'$ step can be implemented by choosing two control qubits from the four candidates.

\begin{table}
\caption{\justifying
Number of qubits and gates required to implement the linear system defined in~\eqref{eq:ConEx} using each method. Resource counts are calculated based on the analyses presented in Section~\ref{sec:CircuitHHL} and Section~\ref{sec:EXamAppl}. Here, $n$ denotes the number of qubits in register~$F$, with the assumption $n \ge 6$. The notation $h(A)$ represents the number of elementary gates required to implement the controlled-$e^{itA}$ operation, which may vary depending on the size of $A$ and the structure of its eigenvalues. Finally, controlled-$U^{\pm 2^{k-1}}$ and controlled-$R_k^{\pm1}$ gates refer to the controlled operations used in both the PE and IPE steps.}
\label{table:compare}
\begin{tabular}{c|c|c|c|c}
\hline
Step& Resource type & Original09 & Hybrid19 & Hybrid25 \\
\hline \hline
\multirow{4}{*}{\makecell[l]{PE$'$, \\ IPE$'$}} & Qubits in $F$ & $n$ & $n$ & 3 \\
 & Hadamard gates & $4n$ & $4n$ & $12$ \\
 & Controlled-$U^{\pm2^{k-1}}$ & $2nh(A)$ & $2nh(A)$ & $6h(A)$ \\
 & Controlled-$R_k^{\pm1}$ & $n^2-n$ & $n^2-n$ & $6$ \\
\hline
CR$'$ & Control qubits & $n$ & $2$ & $2$ \\
\hline
\end{tabular}
\end{table}

The circuit design for Hybrid25, applied to the same example, is presented in Section~\ref{sec:EXamAppl}. As shown in Figure~\ref{fig:example}, the PE$'$ step uses only two distinguishing qubits and a single non-constant qubit following the leading qubit for phase estimation. This significantly reduces the gate overhead. Furthermore, in the CR$'$ step, the two distinguishing qubits serve as control qubits, which is identical to the number used in Hybrid19. However, from the perspective of error accumulation, Hybrid25 offers an advantage: since register $F$ consists of only three qubits, it further mitigates errors originating from qubit decoherence and gate noise. A summary of the qubit and gate counts required by each implementation method is provided in Table~\ref{table:compare}.

\section{Comparison of Circuit Error Rates Between Hybrid19 and Hybrid25} \label{sec:Experiments}

In this section, we implement the HHL algorithm using both Hybrid19 and Hybrid25, and execute each resulting circuit on an actual IBM quantum computer. By comparing the experimental outcomes, we demonstrate that Hybrid25 can effectively reduce the circuit error rate of the HHL algorithm.

Let us consider the same linear system $A\vec{x} = \vec{b}$ introduced in Section~\ref{sec:EXamAppl}. To implement the circuit, we choose $A$ and $\vec{b}$ as follows:
\begin{equation}
A=\begin{bmatrix}
\lambda_1 & 0 & 0 & 0 \\
0 & 0 & 0 & 0 \\
0 & 0 & \lambda_2 & 0 \\
0 & 0 & 0 & \lambda_3
\end{bmatrix},\quad
\vec{b} = \frac{1}{\sqrt{3}}
\begin{bmatrix} 1 \\ 0 \\ 1 \\ 1 \end{bmatrix}= \ket{b}, \label{eq:Ab}
\end{equation}
where the eigenvalues $\lambda_j$ are defined in Eq.~\eqref{eq:eigenvalues}. The solution vector $\vec{x}$ can be obtained by directly solving the linear system. However, the output of the HHL algorithm corresponds to the normalized version of this solution, given by
\begin{equation}
\ket{x}= \frac{1}{\sqrt{\frac{1}{\lambda_1^2}+\frac{1}{\lambda_2^2}+\frac{1}{\lambda_3^2}}}
\begin{bmatrix} \frac{1}{\lambda_1} \\ 0 \\ \frac{1}{\lambda_2} \\ \frac{1}{\lambda_3} \end{bmatrix}
\approx\begin{bmatrix} n_1 \\ n_2 \\ n_3 \\ n_4 \end{bmatrix}
=\begin{bmatrix} 0.5005 \\ 0 \\ 0.8243 \\ 0.2644 \end{bmatrix}. \label{eq:expExam}
\end{equation}

To verify that Hybrid19 and Hybrid25 behave identically to the original HHL algorithm in theory, we first simulate the corresponding circuits using the AerSimulator. For the simulations, we perform $10^6$ shots and measure the qubits in both register $E$ and register $G$. Subsequently, we execute the circuits on IBM's 156-qubit device ibm\_kingston on June 24, 2025, with 4096 shots per run, and measure the qubits in both registers. The processor is classified as a Heron r2 architecture. The Qiskit version used was 2.1.0, and all experiments were carried out in a Jupyter notebook environment via the IBM Quantum platform. In all simulations and experiments presented in this section, the parameter $c = 1/5$ is fixed.

Specifically, in this example, only two qubits are sufficient to represent the input state $\ket{b}$. Accordingly, register $G$ consists of two qubits. The number of qubits in register $F$ is set to the maximum allowed. As the first step of the hybrid method, the QPE algorithm is repeatedly executed to obtain the binary matrix $B$ defined in Eq.~\eqref{eq:concrete}. Based on this matrix, register $F$ consists of six qubits in Hybrid19 and three qubits in Hybrid25. As a result, the total number of qubits required to implement the PE$'$, CR$'$, and IPE$'$ steps is nine for Hybrid19 and six for Hybrid25.

\begin{figure}
\includegraphics[clip,width=.99\columnwidth]{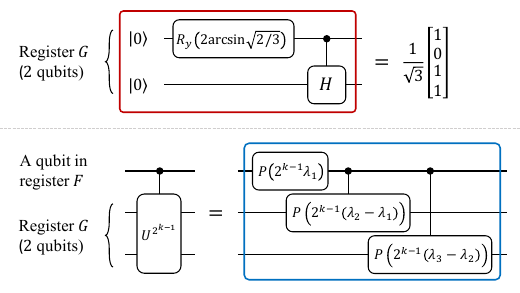}
\caption{\justifying
State preparation and controlled-$U$ construction in hybrid HHL. We consider the linear system $A\vec{x} = \vec{b}$ presented in Eq.~\eqref{eq:Ab}. The upper part of the figure illustrates how the input state $\ket{b}$ is prepared in register $G$ using an $Ry$ gate followed by a controlled-Hadamard gate. The lower part shows how to implement the controlled-$U^{2^{k-1}}$ operation corresponding to matrix $A$ by applying relative phase gates conditioned on a qubit in register $F$.
}
\label{fig:SPnU}
\end{figure}

To prepare the input state $\ket{b}$, an $R_y(2\arcsin(\sqrt{2/3}))$ gate followed by a controlled-Hadamard gate is applied to register $G$, as shown in Figure~\ref{fig:SPnU}. To implement the PE$'$ step, we next apply the controlled-$e^{itA}$ operation, which uses a single control qubit and two target qubits. For the linear system under consideration, the relevant eigenstates are $\ket{00}$, $\ket{10}$, and $\ket{11}$, with $\ket{01}$ excluded. Accordingly, as illustrated in Figure~\ref{fig:SPnU}, a relative phase gate $P(\theta)$ is applied to each eigenstate according to its corresponding eigenvalue, where $P(\theta)$ is defined as
\begin{equation}
P(\theta)
=
\begin{bmatrix}
1 & 0 \\
0 & e^{2\pi i \theta}
\end{bmatrix}.
\end{equation}
The remaining part of the PE$'$ step can be constructed by referring to Section~\ref{sec:IQFT} and Figure~\ref{fig:example}.
To implement the CR$'$ step, both Hybrid19 and Hybrid25 use the CR$'$ circuit shown in Figure~\ref{fig:example}. Finally, the IPE$'$ circuit is constructed as the inverse of the PE$'$ step.

\begin{table}[ht]
\caption{\justifying
Gate counts of transpiled circuits for Hybrid19 and Hybrid25. 
The gate types include single-qubit $\texttt{sx}$ (square-root of $X$), $\texttt{rz}$ (phase rotation about $Z$), $\texttt{x}$ (Pauli-$X$), and the two-qubit $\texttt{cz}$ (controlled-$Z$) gate.
}
\label{table:gateCounts}
\begin{tabular}{l|ccc|c|c}
\hline
Method & \texttt{sx} & \texttt{rz} & \texttt{x} & Single-qubit total & \texttt{cz} \\
\hline \hline
Hybrid19 & 405 & 237 & 8 & 650 & 189 \\
Hybrid25 & 108 & 77 & 8 & 193 & 54 \\
\hline
\end{tabular}
\end{table}

To evaluate the circuit complexity of each method, we compare the number of gates in the transpiled circuits for Hybrid19 and Hybrid25. Table~\ref{table:gateCounts} summarizes the counts of each gate type after transpilation.
Hybrid25 exhibits a significant reduction in gate count compared to Hybrid19, particularly in the number of two-qubit $\texttt{cz}$ gates (54 vs. 189) and single-qubit $\texttt{sx}$ gates (108 vs. 405).

In both the simulations and the hardware experiments, only the measurement outcomes in which register $E$ is observed to be in the state $\ket{1}$ are used to compute the normalized solution.
Theoretically, the expected output distribution should consist of the three bit strings 100, 110, and 111. However, due to noise and gate imperfections in the actual device, an additional bit string, 101, may appear in the results.
The probability associated with each bit string corresponds to the square of the respective component of the normalized solution vector $\ket{x}$. For example, the probability of measuring 100 should approximate $n_1^2$, where $n_1$ is presented in Eq.~\eqref{eq:expExam}.

\begin{figure*}
\centering
\includegraphics[width=\textwidth]{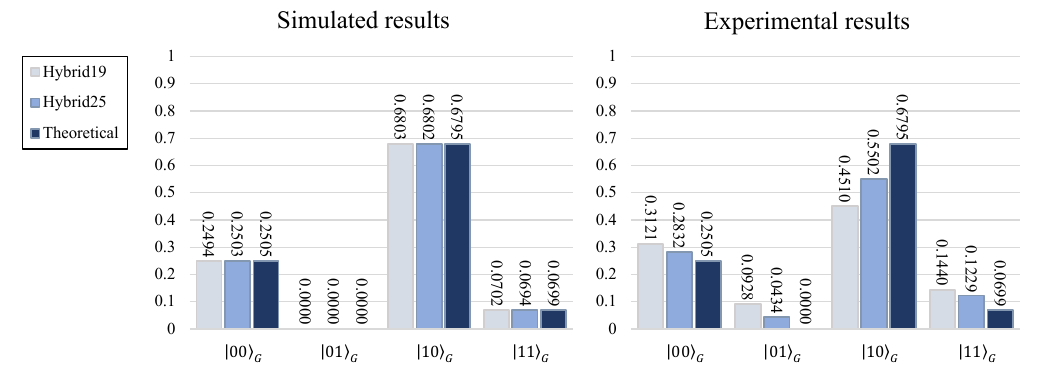}
\caption{\justifying
Comparison of the probability distributions obtained from Hybrid19, Hybrid25, and the theoretical solution.
We implement the HHL algorithm using both Hybrid19 and Hybrid25 for the linear system $A\vec{x} = \vec{b}$ presented in Eq.~\eqref{eq:Ab}, and compare the normalized solutions obtained from the PE$'$, CR$'$, and IPE$'$ circuits.
The horizontal axis represents the basis states of register $G$: $\ket{00}$, $\ket{01}$, $\ket{10}$, and $\ket{11}$. 
The vertical axis shows the squared amplitudes of the components of the normalized solution $\ket{x}$, i.e., the probabilities $n_j^2$.
The left panel shows simulated results obtained using the AerSimulator, where both Hybrid19 and Hybrid25 behave identically to the original HHL algorithm in the absence of noise.
The right panel presents experimental results measured on the ibm\_kingston quantum device.
In this case, Hybrid25 demonstrates a closer match to the theoretical values than Hybrid19, indicating improved circuit accuracy and reduced error rates.
} \label{fig:comparisonResults}
\end{figure*}

Figure~\ref{fig:comparisonResults} shows how closely the results from each simulation and experiment match the ideal probabilities derived from the normalized solution.
As demonstrated in Table~\ref{table:gateCounts}, Hybrid25 implements the circuit with fewer qubits and gates compared to Hybrid19, despite solving the same problem.
This reduction in circuit resources contributes to lower overall circuit error rates and improved execution fidelity on noisy quantum hardware.
These findings demonstrate that the application of QSPE and QPPE can effectively reduce the circuit error rate of the HHL algorithm in practice.

\section{Conclusion}

We have proposed two variations of the QPE algorithm, QSPE and QPPE, and demonstrated their effectiveness in reducing circuit complexity and error rates. By integrating these methods into a hybrid quantum–classical implementation of the HHL algorithm, we achieve significant reductions in both qubit count and gate depth, enhancing the feasibility of such algorithms on near-term quantum hardware.

Given that QPE serves as a central subroutine in a diverse array of quantum algorithms, our techniques hold broad applicability beyond the HHL framework. Notably, they can be adapted to iterative phase estimation protocols~\cite{OLoan2009,Smith2022} and semiclassical Fourier transform~\cite{Griffiths1996}, both of which are favored in resource-constrained environments.

Moreover, since QPE fundamentally relies on the spectral decomposition of Hermitian operators, it is natural to inquire whether our methods can be generalized to algorithms built upon singular value decomposition (SVD)~\cite{Strang2006}. This includes paradigms such as quantum singular value transformation~\cite{Gilyen2019}, quantum principal component analysis~\cite{Lloyd2014}, and SVD-based quantum algorithm~\cite{Bellante2022}. The potential extension of our techniques to these contexts is an intriguing direction for further exploration.

In addition to QPE, the HHL algorithm itself is widely utilized as a subroutine in numerous quantum applications, including quantum differential equation solvers~\cite{Berry2014,Lloyd2020,Liu2021}, quantum machine learning~\cite{Lloyd2013}, big data analysis~\cite{Rebentrost2014} and quantum least‑squares/data‑fitting algorithms~\cite{Wiebe2012}. Therefore, our hybrid method is expected to play a significant role in reducing circuit error rates across these domains. In this work, we introduced concepts such as binary matrix representations and maximal distinguishing column sets to build a theoretical framework for analyzing eigenvalue information in matrix $A$. However, to fully exploit the benefits of our hybrid approach, several open questions remain: (i) It is unclear whether choosing the minimum distinguishing set always leads to optimal performance. (ii) In cases where multiple minimal sets exist, the criteria for selecting among them are not yet established. (iii) Though the original HHL algorithm guarantees exponential speedup under certain conditions, similar rigorous analyses for our hybrid method have not been conducted. We propose these challenges as avenues for future work.

\begin{acknowledgments}
We acknowledge the use of IBM Quantum services for this work. The views expressed are those of the authors, and do not reflect the official policy or position of IBM or the IBM Quantum team.
This research was supported by Basic Science Research Program through the National Research Foundation of Korea (NRF) funded by the Ministry of Education (Grant No. NRF-2020R1I1A1A01058364 and Grant No. RS-2023-00243988).
\end{acknowledgments}


\nocite{*}

\bibliography{vQPE}

\end{document}